# Fully and partially distributed Quantum Generalized Benders Decomposition for Unit Commitment Problems

Fang Gao[1], Dejian Huang[1], Ziwei Zhao[1], Wei Dai[1,*], Mingyu Yang[1], Qing Gao[2] and Yu Pan[3]

[1]School of Electrical Engineering, Guangxi University, Nanning, 530004, China
[2]Hangzhou Innovation Institute of Beihang University, Hangzhou, 310051, China
[3]College of Control Science and Engineering, Zhejiang University, Hangzhou 310027, China

A series of hybrid quantum-classical generalized Benders decomposition (GBD) algorithms are proposed to address unit commitment (UC) problems under centralized, distributed, and partially distributed frameworks. In the centralized approach, the quantum GBD transforms the master problem (MP) into a quadratic unconstrained binary optimization form suitable for quantum computing. For distributed systems, the distributed consensus quantum GBD employs an average consensus strategy to reformulate subproblems into local subproblems. By leveraging the dual information, local cutting planes are constructed to decompose the MP into local master problems (LMPs). This approach reduces the qubit overhead and addresses the partitioning requirements. The consensus-inspired quantum GBD (CIQGBD) and its partially distributed variant, D-CIQGBD are proposed based on optimizing the allocation of relaxation variables directly, the algorithms construct more rational cutting planes, thereby enhancing the minimum eigenenergy gap of the system Hamiltonian during quantum annealing and improving the computational efficiency. Extensive experiments under various UC scenarios validate the performance of the above-mentioned hybrid algorithms. Compared to the classical solver Gurobi, D-CIQGBD demonstrates a speed advantage in solving the security-constrained UC problem on the IEEE-RTS 24-bus system. These results provide new perspectives on leveraging quantum computing for the distributed optimization of power systems.

## 1. Introduction

Quantum computing is recognized as a potent tool for solving optimization problems for addressing classically challenging problems by using quantum superposition and entanglement [1]. Across various fields seeking quantum acceleration, the primary focus is on addressing constrained combinatorial optimization problems [2], whose fundamental challenge stems from the combinatorial growth of discrete variables within the feasible domain space. Traditional methods swiftly encounter computational bottlenecks when confronting this challenge. The expansive search space and parallel computing characteristics have propelled the advancement of quantum computing in addressing constrained combinatorial optimization problems. Quantum algorithms tailored for specific problems, such as the quantum approximate optimization algorithm (QAOA) [3] and quantum annealing, can be implemented on real quantum hardware to address constrained combinational optimization problems based on the Ising model [4], which can be transformed into a quadratic unconstrained binary optimization (QUBO) problem [5], specifically expressed as $\min X^T QX$ , where $X \in \{0,1\}^n$ , $Q \in R^{n \times n}$ . Despite notable

* Corresponding author
*E-mail address:* fgao@gxu.edu.cn (F. Gao), djhuang@st.gxu.edu.cn (D. Huang), 1021313804@qq.com (Z. Zhao), weidai2019@163.com (W. Dai), myyang@st.gxu.edu.cn (M. Yang), gaoqing@buaa.edu.cn (Q. Gao), ypan@zju.edu.cn (Y. Pan)

progress in quantum hardware and algorithms in recent years, the currently accessible quantum hardware is still situated in the era of noisy intermediate-scale quantum (NISQ) [6]. The quantity of available qubits and the error rate of quantum gates are insufficient for effectively tackling practical problems. Meanwhile, developed in the frame of adiabatic quantum computing, quantum annealing incorporates quantum fluctuations to expedite convergence toward the optimal solution and is easier to converge to the global optimal value than classical annealing [7]. Utilizing quantum annealers like D-WAVE significantly expands the scale of QUBO problems that can be tackled. At present, the D-WAVE quantum annealing machine has found practical application in diverse scenarios, including protein structure design [8-10], manufacturing logistics scheduling, path planning [11], and financial portfolio optimization [12,13].

In the domain of power systems, numerous discrete variables associated with logical choices for describing equipment states exist, including decisions like the on/off operations of generator units and the topology of transmission lines. Consequently, constrained combinatorial optimization is prevalent in the field of power systems, with typical examples encompassing the unit commitment (UC) problem. UC is a very significant optimization problem in power system dispatching, falling under the realm of mixed-integer nonlinear programming (MINLP). The UC problem is nonconvex and NP-Hard because of the 0-1 variables. With the proliferation of distributed energy resources (DERs), the computational burden arising from a substantial quantity of binary variables experiences exponential growth, which poses computational challenges to traditional classical techniques. A groundbreaking solving strategy is crucial to effectively solve the UC problem in this context.

Within the framework of the Alternating Direction Method of Multipliers (ADMM), the Quantum ADMM (QADMM) [14] decomposes the UC problem based on variable types. It formulates QUBO subproblems, solvable using QAOA on a quantum computer, and constrained convex subproblems, solvable on a classical computer. Similarly, the Quantum Surrogate Lagrangian Relaxation (QSLR) method [15] decomposes the UC problem within the Surrogate Lagrangian Relaxation (SLR) framework. These two hybrid algorithms rely on quantum circuits and encounter limitations due to the restricted number of available qubits and the shallow depth of quantum circuits in the NISQ era. Contemporary quantum annealers demonstrate the capability to handle larger-scale QUBO problems compared to the currently available quantum circuit-based hardware. Quantum annealing has been employed to solve simplified formulations of the UC problem [16]. Additionally, annealing-based hybrid Benders decomposition has been utilized for multi-energy system optimization [17], as well as for selecting cutting planes of the linearized UC problem [18]. In three other instances [19-21], a hybrid quantum Benders decomposition approach was used to address the mixed-integer linear programming problem. However, this conventional approach to reconstructing the QUBO Hamiltonians of the subproblems in the hybrid algorithms requires the discretization of continuous variables in the subproblems, introducing a substantial number of auxiliary qubits and leading to a considerable increase in the dimension of the Hilbert space, thereby affecting the accuracy of solving QUBO problems and the convergence of the algorithm.

Hen and Sarandy [22] introduced a technique for solving optimization problems by encoding constraints directly and proposed general criteria for constructing driver Hamiltonians corresponding to optimization problem constraints. They demonstrated a notable reduction in the dimensionality of the Hilbert space achieved through constrained quantum annealing and improved the precision of quantum annealing by enlarging the minimum eigenenergy gap between the ground state and the first excited state of the Hamiltonian. Nevertheless, the construction of driver Hamiltonians with multiple commuting constraints heavily depends on specific problem intuitions, and as of now, there is no universally applicable method for generating the corresponding driver Hamiltonian based on arbitrary constraints. Recognizing the essential metric for the adiabatic evolution algorithm is the minimum eigenenergy gap between the ground state and the first excited state of the system Hamiltonian, this research employs consensus-inspired objective functions of relaxed subproblems within the decomposition framework to achieve more concise and rational construction of cutting planes. These improved objective functions and cutting planes increase the minimum eigenenergy gap of the system Hamiltonian, thereby enhancing the accuracy of quantum annealing in solving the QUBO master problem (QUBO-MP). This advancement enables the centralized solution framework to handle larger-scale UC problems effectively.

With the continuous development of smart grids, centralized algorithms struggle to meet the partitioned processing requirements. When applied to the UC problem, a centralized framework consumes substantial computational resources, exhibits low efficiency, and lacks the flexibility to schedule based on local bus load demands. Prior research [23,24] has explored distributed strategies to promote energy transactions among microgrids (MGs). In this research, we employ an average consensus strategy to enable distributed computation by introducing consensus variables in the subproblems, decoupling power balance constraints at the bus level. This approach leads to a fully distributed framework that not only reduces the qubit required for solving the QUBO-MP but also improves computational efficiency.

In summary, this work proposes the quantum generalized Benders decomposition (QGBD) algorithm to address UC problems with minimum on/off constraints, achieving improved computational speed compared to the classical generalized Benders decomposition (GBD) algorithm. To reduce the number of required qubits, a distributed consensus quantum GBD (D-CQGBD) algorithm is proposed, establishing a fully distributed solution framework. To enhance the accuracy of quantum annealing in solving the QUBO-MP, the consensus-inspired quantum GBD (CIQGBD) and its partially distributed variant (D-CIQGBD) are introduced, enabling more rational construction of cutting planes. The main contributions of this work include:

1) This research integrates quantum computing into the GBD framework, proposing a series of hybrid quantum-classical decomposition algorithms. These algorithms transform the NP-hard MP into a QUBO formulation (*i.e.*, QUBO-MP) suitable for quantum computing, enabling efficient solutions using QAOA or quantum annealing. By leveraging the speed advantage of quantum computing in solving QUBO problems, these algorithms accelerate the solution of UC problems. In large-scale UC problems, the hybrid algorithms demonstrate significant speed advantages compared to their classical counterparts.

2) Within the distributed framework, this research proposes the classical D-CGBD algorithm and its hybrid counterpart, D-CQGBD. Both algorithms introduce consensus variables between neighboring MGs to reformulate subproblems into local subproblems within each MG. Using dual information from the local subproblems, local cutting planes are constructed, decomposing the MP into multiple local master problems (LMPs), thus reducing the number of required qubits. The D-CGBD and D-CQGBD methods not only achieve decoupling and acceleration of distributed computation but also ensure the privacy of MG information, meeting the requirements for cross-regional independent decision-making and partitioned management in power grids.

3) This research improves the relaxation variable allocation strategy based on a consensus-inspired approach (CIQGBD), enabling more rational construction of cutting planes. This method enables the MP to be decomposed into multiple LMPs, facilitating distributed parallel processing (D-CIQGBD). The QUBO-form Hamiltonian of the MP, constructed with more rational cutting planes, increases the minimum eigenenergy gap between the ground state and the first excited state of the system Hamiltonian during the annealing process. This enhancement improves the accuracy of quantum annealing in solving the QUBO-MP. In the security-constrained UC (SCUC) problem of the IEEE-RTS-24 bus system, D-CIQGBD demonstrates a speed advantage over the commercial solver Gurobi.

The remaining part of this paper is organized as follows. Section 2 establishes the necessary theoretical background. Section 3 presents the mathematical model for the UC and SCUC problems. Section 4 introduces centralized (QGBD), fully distributed (D-CQGBD), and partially distributed quantum decomposition (D-CIQGBD) algorithms. Section 5 presents the results and corresponding discussions. Section 6 concludes the work.

## 2. Preliminaries

This section introduces the QAOA [3] and quantum annealing algorithms [25], both of which are applicable to combinatorial optimization.

### 2.1. QAOA

QAOA is a gate-based quantum optimization algorithm, which has a potential advantage in combinatorial optimization problems. By optimizing the parameters of the quantum gates (*i.e.*, $\vec{\gamma},\vec{\lambda} \in \boldsymbol{R}^m$), QAOA aims to minimize the expectation value of the problem Hamiltonian $H_p$ :

$$E_p\left(\vec{\gamma},\vec{\lambda}\right)=\left\langle\varphi\left(\vec{\gamma},\vec{\lambda}\right)\middle|H_p\middle|\varphi\left(\vec{\gamma},\vec{\lambda}\right)\right\rangle. \tag{1}$$

When $E_p\left(\vec{\gamma},\vec{\lambda}\right)$ is minimized, we perform measurement on $\left|\varphi\left(\vec{\gamma},\vec{\lambda}\right)\right\rangle$ to obtain the solution of the QUBO problem.

As shown in **Figure 1**, in the QAOA algorithm, the initial state before entering Layer 1 is an equal probability superposition of all states, which is realized by a series of Hadmard gates. The initial state can be expressed as:

$$|\varphi_0\rangle=\frac{1}{\sqrt{2^n}}\sum_{q\in\{0,1\}^n}|q\rangle, \tag{2}$$

where $n$ is the number of binary variables of the QUBO problem. In each layer, the state is first rotated along $H_{\mathrm{p}}$ with the angle $\lambda_i$, and then rotated along the transverse Hamiltonian $H_{\mathrm{b}}$ with the angle $\gamma_i$. Here $H_{\mathrm{b}}$ is taken to be the Pauli-X operator:

$$H_{\mathrm{b}}=\hat{\sigma}_x=\begin{bmatrix}0 & 1\\ 1 & 0\end{bmatrix}, \tag{3}$$

which leads to the evolution operator in each layer as follows:

$$U\left(H_b,\gamma_i\right)=\left(e^{\frac{-i\hat{\sigma}_x\gamma_i}{2}}\right)^{\otimes n}=\left(\cos\left(\frac{\gamma_i}{2}\right)I-i\sin\left(\frac{\gamma_i}{2}\right)\hat{\sigma}_x\right)^{\otimes n}=\begin{bmatrix}\cos\left(\frac{\gamma_i}{2}\right) & -i\cdot\sin\left(\frac{\gamma_i}{2}\right)\\ -i\cdot\sin\left(\frac{\gamma_i}{2}\right) & \cos\left(\frac{\gamma_i}{2}\right)\end{bmatrix}^{\otimes n}=\left(R_x\left(\gamma_i\right)\right)^{\otimes n}. \tag{4}$$

The evolution operator corresponding to $H_p$ in each layer is $U\left(H_p,\lambda_i\right)=e^{-iH_p\lambda_i}$ .

After $m$ layers, the quantum state evolves to:

$$\left|\varphi\left(\vec{\gamma},\vec{\lambda}\right)\right\rangle=\prod_{i=1}^{m}U\left(H_b,\gamma_i\right)U\left(H_p,\lambda_i\right)\cdot|\varphi_0\rangle. \tag{5}$$

Theoretically, the minimum of $E_p$ can be approached when the number of layers is infinitely large:

$$\lim_{m\to\infty}E_p\left(\vec{\gamma}^*,\vec{\lambda}^*\right)=\left\langle\varphi\left(\vec{\gamma}^*,\vec{\lambda}^*\right)\middle|H_p\middle|\varphi\left(\vec{\gamma}^*,\vec{\lambda}^*\right)\right\rangle=E_{\min}. \tag{6}$$

Given a finite number of layers in practice, $E_p$ is minimized by iteratively optimizing the values of the parameters $\lambda_i$ and $\gamma_i$ with one classical optimizer. Labelling $C_k$ as the minimum of $E_p$ for $k$ iterations, the optimization converges at the $k$-th iteration when the following condition is satisfied:

$$|C_k - C_{k-1}| < \varepsilon . \tag{7}$$

Here $\varepsilon$ is the tolerance error.

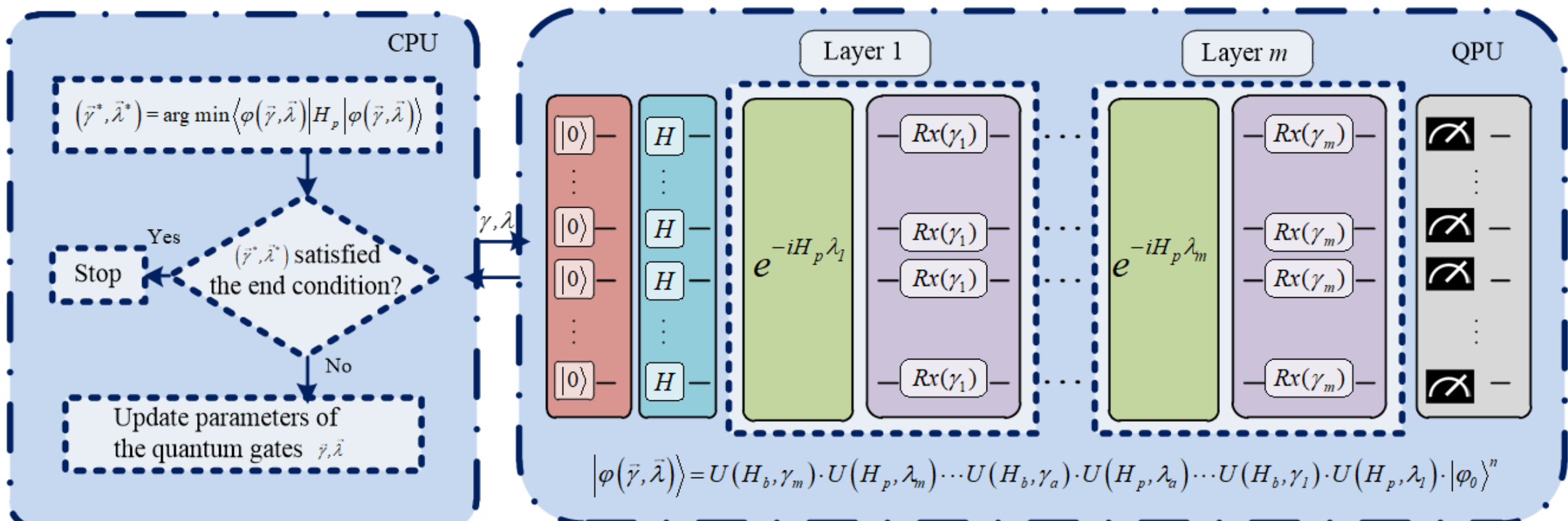

**Figure 1.** The framework of QAOA algorithm.

### 2.2. Quantum Annealing

Quantum annealing belongs to quantum adiabatic evolution. During this evolution, the system stays at its lowest energy state. At the end of quantum annealing, the quantum state is just the ground state of the problem Hamiltonian.

The quantum annealing algorithm can be implemented with D-WAVE. The Hamiltonian of the D-WAVE quantum annealing machine can be expressed with the Ising model:

$$H_{Ising} = -\frac{A(s)}{2} H_b + \frac{B(s)}{2} H_p = -\frac{A(s)}{2}\left(\sum_i \hat{\sigma}_x^{(i)}\right) + \frac{B(s)}{2}\left(\sum_i h_i \hat{\sigma}_z^{(i)} + \sum_{(i,j)} k_{i,j} \hat{\sigma}_z^{(i)} \hat{\sigma}_z^{(j)}\right), \tag{8}$$

where $s \in [0,1]$ is the normalized anneal fraction. $H_b$ is the driver Hamiltonian, while $H_p$ is the problem Hamiltonian, whose ground state is just the solution of the QUBO model.

With almost no interaction with the external environment, the quantum system can evolve slowly enough from the initial Hamiltonian $H_b$ to the problem Hamiltonian $H_p$:

$$H(\tau) = \left(1 - \frac{\tau}{\mathcal{T}}\right) H_b + \frac{\tau}{\mathcal{T}} H_p, \quad \frac{\tau}{\mathcal{T}} \in [0,1], \tag{9}$$

where $\tau$ is the timing point, and $\mathcal{T}$ is the whole annealing time.

The principle of the quantum annealing algorithm can be described as follows: (1) At $\frac{\tau}{\mathcal{T}} = 0$, the quantum system is in the ground state of the initial Hamiltonian $H_b$, that is, all qubits are in the superposition of $|0\rangle$ and $|1\rangle$ (*i.e.*, $\frac{1}{\sqrt{2}}(|0\rangle + |1\rangle)$); (2) When the quantum annealing algorithm starts ($0 < \frac{\tau}{\mathcal{T}} < 1$), the quantum system is slowly perturbed, and couplings $k_{i,j}$ and biases $h_i$ in the problem Hamiltonian $H_p$ are introduced, making the qubits become entangled. Assuming the perturbation is long enough and changes slowly enough (*i.e.*, the ideal quantum annealing process), the quantum system will remain in the ground state without transitioning to the excited states; (3) When $\frac{\tau}{\mathcal{T}} = 1$, the quantum annealing process ends. Because the quantum system does not have an energy level crossover during the whole quantum annealing process, the system is now in the ground state of the problem Hamiltonian, which can be measured to give the solution of the QUBO model.

## 3. Mathematical Model of the UC and SCUC Problems

The UC and SCUC problems [26-27] is to minimize the operation cost of the power system $\mathcal{J}(U)+\mathcal{K}(P)$ (including the total fuel cost and the commitment cost) by adjusting the on/off status *U* and the output power *P* of each unit under a series of system constraints (e.g., system power demand constraints, generator output power constraints, generator minimum on/off time constraints):

$$\min_{P,U} \mathcal{J}(U)+\mathcal{K}(P), \tag{10a}$$

$$\mathcal{J}(U)=\sum_{i=1}^{I}\sum_{t=1}^{T}\mathcal{J}_{i,t}(U)=\sum_{i=1}^{I}\sum_{t=1}^{T} d_i \cdot u_{i,t}, \tag{10b}$$

$$\mathcal{K}(P)=\sum_{i=1}^{I}\sum_{t=1}^{T}\mathcal{K}_{i,t}(P)=\sum_{i=1}^{I}\sum_{t=1}^{T} a_i \cdot p_{i,t}^2 + b_i \cdot p_{i,t} + c_i, \tag{10c}$$

where *I* and *T* represent the total number of units and time periods, respectively. $U \in \{0,1\}^{IT\times 1}$ and $P \in \boldsymbol{R}^{IT\times 1}$ are vectors composed of decision variables $u_{i,t}$ and $p_{i,t}$. Variable $p_{i,t}$ represents the output power of the unit *i* on the time period *t*, and $u_{i,t}$ represents the on/off status of the unit *i* on the time period *t* (0 indicates that the unit is "off", while 1 indicates that the unit is "on"). $a_i, b_i, c_i$ are the fuel cost parameters of the unit *i*, and $d_i$ indicates the startup/shutdown cost of the unit *i*.

The UC problem is subject to the following constraints:

$$\text{s.t.} \sum_{i=1}^{I} p_{i,t} = D_t, \quad \forall t \tag{11a}$$

$$p_i^{min} \cdot u_{i,t} \le p_{i,t} \le p_i^{max} \cdot u_{i,t}, \quad \forall i, \forall t \tag{11b}$$

$$\sum_{\tau=t-T_i^{on}}^{t-1} u_{i,\tau} \ge T_i^{on} \cdot u_{i,t-1}\left(1-u_{i,t}\right), \quad \forall i, \forall t \ge T_i^{on} \tag{11c}$$

$$\sum_{\tau=t-T_i^{off}}^{t-1} \left(1-u_{i,\tau}\right) \ge T_i^{off} \cdot u_{i,t}\left(1-u_{i,t-1}\right), \quad \forall i, \forall t \ge T_i^{off} \tag{11d}$$

Here $D_t$ represents the system load demand at time period *t*; The parameters $p_i^{\min}$ and $p_i^{\max}$ denote the minimum and maximum output power of the unit *i*, respectively, while $T_i^{on}$ and $T_i^{off}$ represent the minimum on and off durations of unit *i*, respectively. The system power demand constraints (11a) stipulate that the total unit output power in time period *t* equals the system load demand for that period. The minimum and maximum unit output power constraints (11b) delineate the maximum and minimum limits on the unit output power. The unit minimum on/off time constraints (11c) and (11d) impose restrictions that the unit maintains its on/off status for a specified duration. Building upon these, the SCUC problem refines the system power demand constraints (11a) into individual power balance constraints for each bus:

$$\sum_{i\in NG(n)} P_{i,t} - \sum_{l\in NL_{from}(n)} P_{l,t} + \sum_{l\in NL_{to}(n)} P_{l,t} = D_{n,t}, \forall n, \forall t \tag{11e}$$

Here $NL_{from}(n)$ and $NL_{to}(n)$ represent the sets of outgoing and incoming transmission lines at bus *n*; $P_{l,t}$ represents the power flow on transmission line *l* at time period *t*; $D_{n,t}$ represents the load demand of bus *n* at time period *t*. The SCUC problem also considers the system power flow constraints and the transmission line power flow capacity constraints:

$$\left(\theta_{n,t}-\theta_{m,t}\right)/x_l - P_{l,t} = 0, \quad \forall l(n\to m), \forall t \tag{11f}$$

$$-P_l^{\max} \cdot Z_{l,t} \le P_{l,t} \le P_l^{\max} \cdot Z_{l,t}. \quad \forall l, \forall t \tag{11g}$$

Here $\theta_{n,t}$ denotes the phase angle at bus *n* at time period *t*; The parameter $P_l^{\max}$ represents the maximum power flow limit of transmission line *l*; The power flow constraint (11f) regulates the distribution of power flow across transmission lines and constrains the phase angle differences between adjacent buses. Constraint (11g) imposes limits on the capacity of power flows through transmission lines. For convenience and without loss of generality, the power flow constraint (11e) is not considered in the formula derivation. The constraints (11) can be summarized into the following three concise forms, which will be referenced in the subsequent content:

$$A \cdot U \leq B, \tag{12a}$$

$$C \cdot P \leq D, \tag{12b}$$

$$E \cdot U + F \cdot P \leq G. \tag{12c}$$

Here $A \in \boldsymbol{R}^{O \times IT}$, $B \in \boldsymbol{R}^{O \times 1}$, $C \in \boldsymbol{R}^{l \times IT}$, $D \in \boldsymbol{R}^{l \times 1}$, $E \in \boldsymbol{R}^{W \times IT}$, $F \in \boldsymbol{R}^{W \times IT}$ and $G \in \boldsymbol{R}^{W \times 1}$. Constraint (12a) is exclusively related to binary variables $u_{i,t}$ and corresponds to the generator minimum on/off time constraints (11c) and (11d). Constraint (12b) is exclusively related to continuous variables $p_{i,t}$ and corresponds to the system power demand constraints (11a). Constraint (12c) involves both binary and continuous variables and corresponds to the generator output power constraints (11b).

## 4. Quantum-empowered Benders Decomposition Algorithms

GBD can be used to deal with MINLP problems. Quantum algorithms (*i.e.,* QAOA and quantum annealing) have advantages in solving combinatorial optimization problems expressed in the QUBO form. Taking advantage of both GBD and quantum algorithms, a quantum decomposition algorithm named QGBD is proposed. In the context of distributed UC problems, we introduce consensus among different MGs and propose two fully distributed consensus algorithms, D-CGBD and its quantum version D-CQGBD. Both the LMPs and local sub-problems are formulated on a per MG basis. Without disclosing the private information within each MG, the algorithms achieve parallel solutions by reaching a consensus on output power for each MG. Building upon this, we propose consensus-inspired partially distributed algorithms, D-CIGBD and its quantum version D-CIQGBD. The sub-problems, consisting of only simple variables (continuous variables in the UC problem), are solved centrally. While the LMPs, which mainly contain complex variables (binary variables in the UC problem) are solved in a parallel and distributed manner. This approach allows for efficient solutions by focusing on different types of variables in a coordinated manner.

The convergence of the Benders decomposition algorithm [28] and its variants has been demonstrated in research [29-30]. Building upon the GBD framework and establishing specific convergence criteria for upper and lower bounds, the aforementioned algorithms are proposed to theoretically converge to optimal solutions that comply with the established convergence criteria within finite steps. In the decomposition framework, continuously constructed feasibility and optimality cutting planes are utilized. Feasibility cutting planes exclude binary variable solutions that fail to meet the requirements, while optimality cutting planes reveal the optimality of a certain binary variable solution under the termination criteria. By iteratively adding the cutting planes, the algorithms narrow down the search space of the binary variables, ensuring that only valid solutions are considered, thus helping identify the best feasible solutions as the algorithms progress. The corresponding quantum versions of the algorithms accelerate the solution of MP/LMP without affecting the inherent property of the finite-step convergence within the decomposition framework.

The integration of quantum computing allows for more efficient processing of large-scale MPs/LMPs, revealing that the combination of the GBD decomposition framework with quantum computing offers an effective approach to solving the UC problem. In the following, Subsections 4.1, 4.2 and 4.3 give the overall frameworks of QGBD, D-CQGBD, and D-CIQGBD, respectively. Subsection 4.4 describes how to transform the MP into the QUBO and Ising form amenable for quantum computation.

### 4.1. Framework of QGBD

QGBD can solve the UC problem in an iterative way. In the $\mathcal{E}$-th iteration, the UC problem can be modeled by the MINLP problem, with Eq. (10) as its objective to be optimized and Eq. (12) as its constraints. After fixing binary variables $U$ as $U^*$, the UC problem is transformed into the following nonlinear optimal economic programming sub-problem, which can be handled by the classical optimizer to give the upper bound (UB) of the original problem.

$$\overline{Z}^{\mathcal{E}} = \min_{P} Z^{\mathcal{E}} = \mathcal{J}\left(U^*\right) + \mathcal{K}\left(P\right) \tag{13a}$$

$$\text{s.t. } C \cdot P \le D, \tag{13b}$$

$$G - F \cdot P \ge E \cdot U^*. \tag{13c}$$

---

**Algorithm 1:** QGBD

---

**Input:** Maximum iteration: $Iter_{\max}$, cost coefficients: $a_i, b_i, g, d_i$, constraints coefficients: $A, B, C, D, E, F, G$;

1. $UB = +\infty, LB = -\infty$, $\mathcal{E} = 1$, $U^* = random$; // Initialize.
2. Set error tolerance $\omega$;
3. **while** $UB - LB \ge \omega$ and $\mathcal{E} \le Iter_{\max}$ **do**:
4. $U$ in the UC problem is fixed as $U^*$ to obtain the sub-problems;
5. **if** the sub-problem is feasible:
6. Solve the sub-problem with $U = U^*$;
7. Obtain the dual information $L_{\mathcal{E}}^{op}, M_{\mathcal{E}}^{op}$ and continue variables solution $P_{\mathcal{E}}^{op}$;
8. Obtain the objective function $\overline{Z}^{\mathcal{E}}$;
9. **if** $\overline{Z}^{\mathcal{E}} \le UB$:
10. update $UB = \overline{Z}^{\mathcal{E}}$;
11. **end**
12. Construct optimality Benders cutting planes using the dual information $L_{\mathcal{E}}^{op}, M_{\mathcal{E}}^{op}$
13. $Z^{\mathcal{E}} \ge \mathcal{J}\left(U\right) + \mathcal{K}\left(P_e^{op}\right) + \left(L_e^{op}\right)^T \left(C \cdot P_e^{op} - D\right) + \left(M_e^{op}\right)^T \left(E \cdot U + F \cdot P_e^{op} - G\right), \forall e \in [1, \mathcal{E}]$
14. Return optimality Benders cutting planes to MP;
15. **else:**
16. Introduce relaxation variables $S \in \left\{ {}^1S, {}^2S \right\}$;
17. Construct and solve the feasible sub-problems;
18. Obtain the dual information $L_{\mathcal{E}}^{fea}, M_{\mathcal{E}}^{fea}$;
19. Construct feasibility Benders cutting planes using the dual information $L_{\mathcal{E}}^{fea}, M_{\mathcal{E}}^{fea}$
20. $0 \ge \left(L_e^{fea}\right)^T \left(C \cdot P_e^{fea} - D\right) + \left(M_e^{fea}\right)^T \left(E \cdot U + F \cdot P_e^{fea} - G\right). \forall e \in [1, \mathcal{E}]$;
21. Decouple the feasibility Benders cutting planes on the time separations;
22. Return multiple feasibility Benders cutting planes;
23. **end**
24. Introduce auxiliary binary variables $S^{cons}, S^{fea}$, penalties $\xi^{cons}, \xi^{fea}$;
25. Transform MP into QUBO-MP $H$;
26. Transform QUBO-MP $H$ into the problem Hamiltonian $H_p$;
27. Map the Ising model to quantum annealing devices;
28. Solve QUBO-MP with quantum annealing to obtain solution $U_{\mathcal{E}}$;
29. Set $\underline{Z}^{\mathcal{E}} = \min \left\{ c_e^{op} + \left(f_e^{op}\right)^T \cdot U_{\mathcal{E}} \middle| e = 1, \ldots, \mathcal{E} \right\}$;
30. **if** $\underline{Z}^{\mathcal{E}} \ge LB$:
31. update $LB = \underline{Z}^{\mathcal{E}}$;
32. **end**
33. Update $U^* = U_{\mathcal{E}}$;
34. $\mathcal{E} = \mathcal{E} + 1$;
35. **end while**;

**Output:** Hourly on/off states $U = U^*$, output power $P = P_{\mathcal{E}}^{op}$ of the units, the operation charges $UB$.

---

When the above sub-problem has a feasible solution, the corresponding Lagrange multipliers $L^{op}$ and $M^{op}$, together with $P^{op}$ up to the $\mathcal{E}$-th iteration, give the optimality Benders cutting planes, which are added as Eq. (15c) in the MP to modify the optimization region. Otherwise, the sum deviation of Eqs.(13b) and (13c) is minimized in the resulted feasible sub-problem:

$$\min_{P} (\| {}^{1}S \|_{1} + \| {}^{2}S \|_{1}) \tag{14a}$$

$$s.t. \quad C \cdot P \leq D + {}^{1}S, \tag{14b}$$

$$G - F \cdot P + {}^{2}S \geq E \cdot U^{*}. \tag{14c}$$

$$ {}^{1}S, {}^{2}S \geq 0. \tag{14d}$$

The corresponding Lagrange multipliers $L^{fea}$ and $M^{fea}$ as well as $P^{fea}$ up to the $\mathcal{E}$-th iteration give the feasibility cutting planes, which are added as Eq. (15d) in the MP to modify the optimization region. Then the following MP can be constructed to give the lower bound (LB) of the original problem:

$$\underline{Z}^{\mathcal{E}} = \min_{U} Z^{\mathcal{E}} \tag{15a}$$

$$s.t. \quad A \cdot U \leq B, \tag{15b}$$

$$Z^{\mathcal{E}} \geq \mathcal{J}(U) + \mathcal{K}\left(P_e^{op}\right) + \left(L_e^{op}\right)^T \left(C \cdot P_e^{op} - D\right) + \left(M_e^{op}\right)^T \left(E \cdot U + F \cdot P_e^{op} - G\right), \forall e \in [1, \mathcal{E}] \tag{15c}$$

$$0 \geq \left(L_e^{fea}\right)^T \left(C \cdot P_e^{fea} - D\right) + \left(M_e^{fea}\right)^T \left(E \cdot U + F \cdot P_e^{fea} - G\right), \forall e \in [1, \mathcal{E}] \tag{15d}$$

Constraints (12b) and (12c) considered in the MP are usually decouplable on different time periods. Therefore, each feasibility cutting plane in the *e*-th iteration (*i.e.*, Eq. (15d)) can be split into *T* feasibility cutting planes:

$$0 \geq \left(L_{e,t}^{fea}\right)^T \left(C_t \cdot P_{e,,t}^{fea} - D_t\right) + \left(M_{e,t}^{fea}\right)^T \left(E_t \cdot U_t + F_t \cdot P_{e,t}^{fea} - G_t\right). \forall e \in [1, \mathcal{E}], \quad \forall t \in [1, T] \tag{15e}$$

In this way, QGBD can obtain enough feasibility cutting planes in fewer iterations, such that ensures a feasible solution of the sub-problem by adding the Benders feasibility cutting planes to the MP more efficiently.

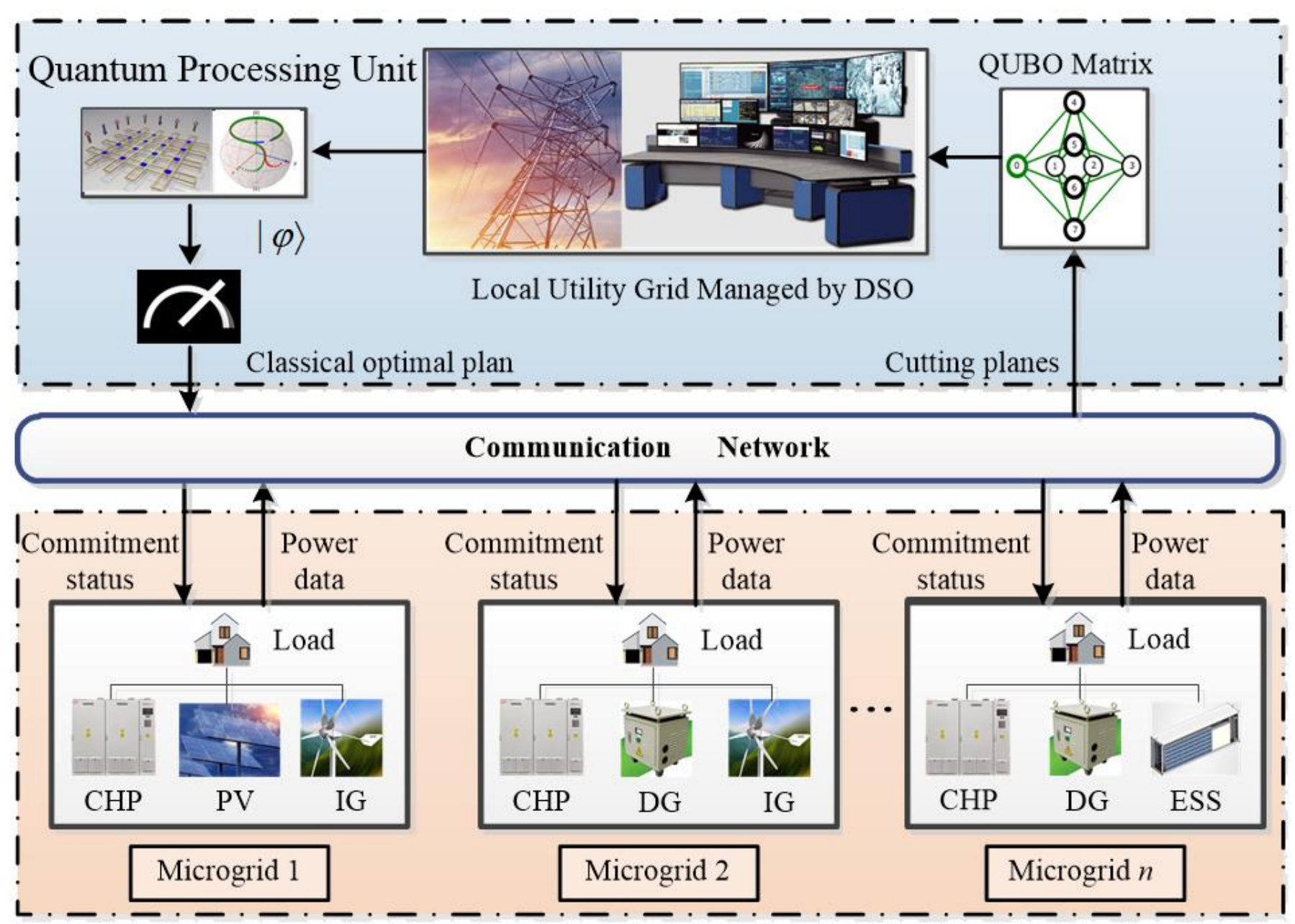

**Figure 2.** Centralized coordination architecture of QGBD.

The procedures of QGBD are described in **Algorithm 1**. The classical Gurobi9 nonlinear solver is used to solve the optimal economic programming sub-problem and relaxed feasible sub-problem with only continuous variables, while the MP is dealt with by the quantum computing step, which is marked in yellow. The quantum computing step transforms the MP (15) into the QUBO-MP as elucidated in Subsection 4.4, which can be solved by QAOA or quantum annealing algorithm efficiently. The solution of the QUBO-MP $U_{\mathcal{E}}$ is fed into the sub-problem as $U^{*}$ for the next iteration, and the total number of iterations $\mathcal{E}$ is increased by 1. The UB and LB are, respectively, updated when $\bar{Z}^{\varepsilon} \leq UB$ and $\underline{Z}^{\varepsilon} \geq LB$ . QGBD ends when the difference between the UB and LB converges to the given threshold, and the converged bound is regarded as the minimized operation cost in the UC problem.

**Figure 2** depicts the architecture of QGBD. The quantum simulator is employed to solve the QUBO-MP. When the quantum simulation ends, the UB of the operation cost of the power system as well as the classical optimal plan (*i.e.,* the unit on/off status in the MGs) is obtained by measurement. After each MG receives the on/off status sent by the distribution system operator (DSO), the unit output powers are obtained by solving the sub-problems. MGs return the Benders cutting planes to the DSO. QGBD converts the Benders cutting planes and generator minimum on/off time constraints into the QUBO matrix, which is provided to the DSO for mapping to the quantum simulator, whose measurement results provide the optimal commitment plan in the next iteration.

### 4.2. Framework of D-CQGBD

In this subsection, we introduce the consensus among various MGs and propose the fully distributed algorithm D-CQGBD. As depicted in **Figure 3**, the entire power system is divided into individual MGs, each comprising a quantum processor and an economic dispatch coordinator (classical optimizer). Introducing the concept of power flow as a consensus variable among different MGs, D-CQGBD iterates through multiple steps to reach an agreement on the power flow between the MGs. This consensus on the power flow ensures that all MGs satisfy the system power demand constraints. Within each MG, the LMP and local sub-problem are solved through distributed parallel computation using their respective quantum processor and classical optimizer. The quantum processor determines the on/off status of the units within the MG. Meanwhile, the classical optimizer is in charge of obtaining the optimal output power or optimal relaxed power under the current state. This information is then utilized to construct cutting planes, which take the form of constraints on the LMP to achieve optimization. The distributed and parallel approach ensures that each MG independently contributes to the overall solution, leading to an effective and scalable optimization process for the power system.

The difference between D-CQGBD and QGDB lies in the incorporation of power flow as a consensus variable among different MGs. The entire solving process of D-CQGBD is distributed and parallel, with each MG being treated as an individual. The ability to interact and cooperate among MGs ensures a more practical and scalable solution for UC problems in real power systems. For each MG *n*, *m* represents a neighboring MG. In the sub-problem, the local variable $\vec{P}_{n,m}^{n}$ is introduced to represent a local copy within MG *n* of the power flow between MG *n* and MG *m*. This local variable corresponds to the global variable $\vec{P}_{n,m}^{+}$, which represents the power flow between MGs *n* and *m*. The consensus constraints $\vec{P}_{n,m}^{n} - \vec{P}_{n,m}^{+} = 0$ drive the local variable $\vec{P}_{n,m}^{n}$ to converge towards the global variable $\vec{P}_{n,m}^{+}$. These consensus constraints are treated as soft constraints and are applied to the objective function of the consensus sub-problem, resulting in the construction of the augmented Lagrange function form of the objective function as:

$$\min_{P_n} \mathcal{J}\left(U_n^*\right) + \mathcal{K}\left(P_n\right) + \sum_{m \in \mathcal{N}_n^{ad}} \lambda_{n,m}^{n} \left(\vec{P}_{n,m}^{n} - \vec{P}_{n,m}^{+}\right) + \frac{\rho_{n,m}^{n}}{2} \left(\vec{P}_{n,m}^{n} - \vec{P}_{n,m}^{+}\right)^2 \tag{16a}$$

$$s.t. \quad C_n \cdot P_n + \vec{P}_{n,m}^{n} + \vec{P}_{n,m}^{n} \leq D_n, \tag{16b}$$

$$G_n - F_n \cdot P_n \geq E_n \cdot U_n^*. \tag{16c}$$

Here $\lambda_{n,m}^{n}$ and $\rho_{n,m}^{n}$ represent the dual variables and penalty coefficients associated with the consensus constraints. The subscripts (*i.e.*, *n,m*) and superscripts (*i.e.*, *n*) together indicate the corresponding copy within MG *n* of the consensus constraints concerning the interaction between MG $n$ and MG $m$. Similar to QGBD, the consensus local sub-problem is a nonlinear economic dispatch sub-problem with fixed binary variables. It can be handled by classical optimizers to obtain the local upper bounds of generation costs for the units within the MG. The consensus local sub-problem solely consists of continuous variables related to the units within the MG and local consensus variables associated with neighboring MGs. Compared to the sub-problems in QGBD, it has lower complexity, a smaller problem size, and can be solved in parallel.

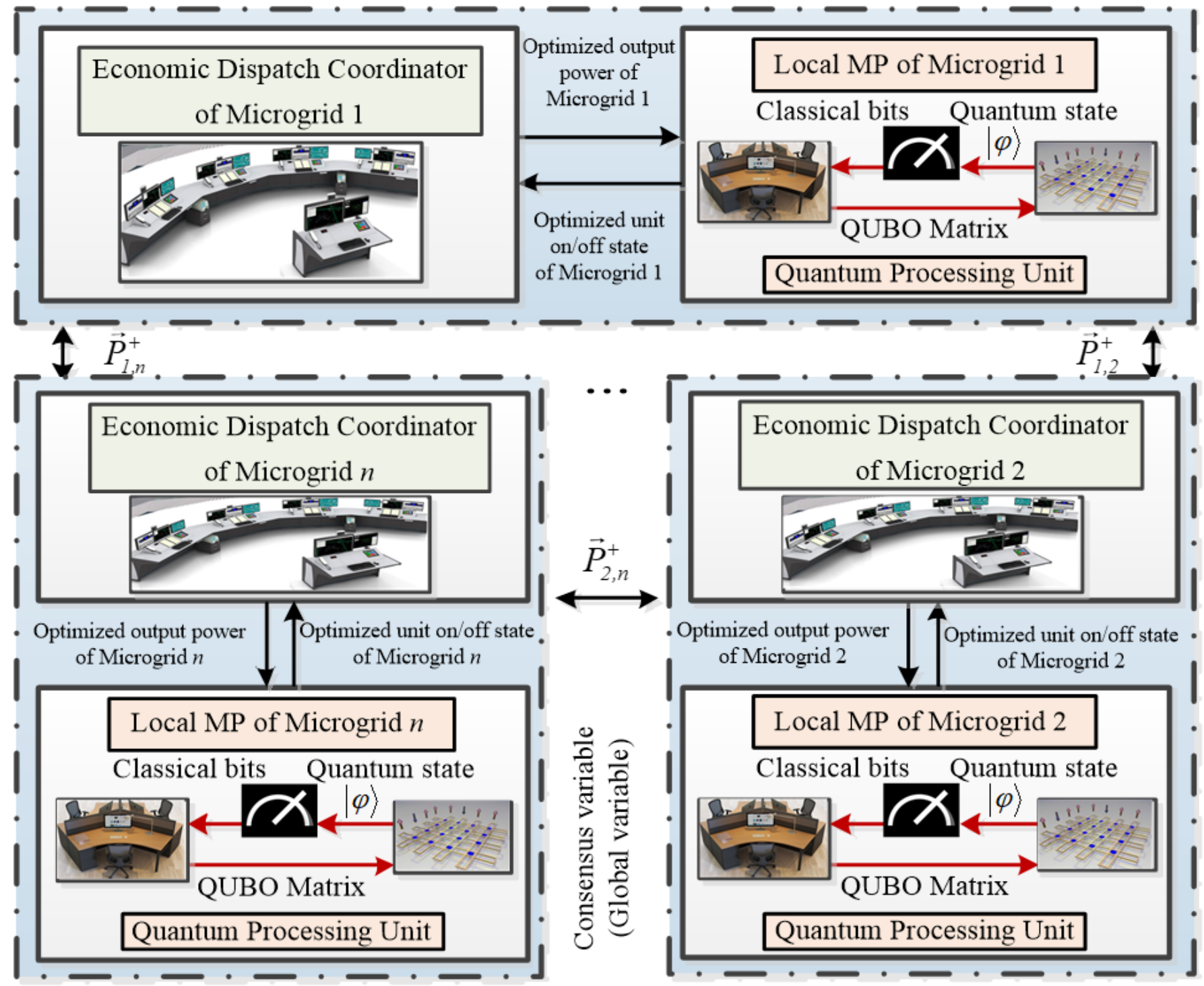

**Figure 3.** Distributed coordination architecture of D-CQGBD.

When the consensus local sub-problems have feasible solutions, a series of iterations are performed to make the consensus among different neighboring MGs converge, that is, the local variables within two neighboring MGs approach the corresponding global variables. The process is formulated in Eq. (17) and Eq. (18) using the $k$-th iteration as an example. An average consensus approach is employed to update the global variables. The dual variables associated with the consensus constraints are updated according to the difference between the local and global variables during each consensus iteration. The residual of the consensus variables serves as the stopping criterion for the consensus updating process.

$$\vec{P}^{+}_{n,m}(k+1) = \frac{\vec{P}^{n}_{n,m}(k) + \vec{P}^{m}_{n,m}(k)}{2} \tag{17a}$$

$$\lambda^{n}_{n,m}(k+1) = \lambda^{n}_{n,m}(k) + \rho^{n}_{n,m} \cdot \left(\vec{P}^{n}_{n,m}(k) - \vec{P}^{+}_{n,m}(k)\right), \tag{17b}$$

$$r^{n}_{n,m}(k) = \left|\vec{P}^{n}_{n,m}(k) - \vec{P}^{+}_{n,m}(k)\right|. \tag{17c}$$

Consensus iterations can be ended when the residual of the consensus variables (17c) falls below a predefined threshold, indicating that the consensus among MGs has reached a satisfactory level and the global optimal value of the sub-problem has been attained. In the final round of consensus iterations, the Lagrange multipliers $L^{op}_{n}$ and $M^{op}_{n}$ corresponding to constraints (16b) and (16c) are obtained within MG $n$, along with the output power $P^{op}_{n}$ of the units and the power flow $\vec{P}^{n}_{n,m}$ associated with the neighboring MG $m$. With this information, the local optimality cutting plane (19c) for the MG $n$ can be derived, which is then incorporated into the LMP to refine the solution space for the local binary variables.

When the local sub-problem has no feasible solution, similar to the feasible sub-problem in QGBD, local relaxation variables are introduced, resulting in the consensus local feasible sub-problem:

$$\min_{P_n} \mathcal{J}\left(U_n^*\right)+\mathcal{K}\left(P_n\right)+\sum_{m\in\mathcal{N}_n^{ad}} \lambda_{n,m}^n\left(\vec{P}_{n,m}^n-\vec{P}_{n,m}^+\right)+\frac{\rho_{n,m}^n}{2}\left(\vec{P}_{n,m}^n-\vec{P}_{n,m}^+\right)^2 +\| {}^1S \|_1 +\| {}^2S \|_1 \tag{18a}$$

$$s.t. \quad C_n \cdot P_n + \vec{P}_{n,m}^n + \vec{P}_{n,m}^n - {}^1S_n \le D_n, \tag{18b}$$

$$G_n - F_n \cdot P_n + {}^2S_n \ge E_n \cdot U_n^*. \tag{18c}$$

$${}^1S_n, {}^2S_n \ge 0. \tag{18d}$$

Similar to how the consensus local sub-problem is handled, the neighboring MG $n$ and MG $m$ reach a consensus through multiple consensus iterations, resulting in the determination of the corresponding Lagrange multipliers $L_n^{fea}$ and $M_n^{fea}$, along with the relaxed output power $P_n^{fea}$ and consensus variable $\vec{P}_{n,m}^n$ within MG $n$. Subsequently, the local feasibility cutting plane is generated from this information and added to the LMP as (19d) to modify the optimization region. The following LMP can be constructed to provide a local lower bound for the generation cost of the units within the MG:

$$\underline{Z}_n^{\varepsilon} = \min_{U_n} Z_n^{\varepsilon} \tag{19a}$$

$$s.t. \quad A_n \cdot U_n \le B_n, \tag{19b}$$

$$Z_n^{\mathcal{E}} \ge \mathcal{J}\left(U_n\right)+\mathcal{K}\left(P_{e,n}^{op}\right)+\left(L_{e.n}^{op}\right)^T\left(C_n\cdot P_{e,n}^{op}-D_n\right)+\left(M_{e,n}^{op}\right)^T\left(E_n\cdot U_n+F_n\cdot P_{e,n}^{op}-G_n\right), \forall e\in[1,\mathcal{E}] \tag{19c}$$

$$0\ge\left(L_{e,t,n}^{fea}\right)^T\left(C_n\cdot P_{e,t,n}^{fea}-D_n\right)+\left(M_{e,t,n}^{fea}\right)^T\left(E_n\cdot U_{t,n}+F_n\cdot P_{e,t,n}^{fea}-G_n\right), \forall e\in[1,\mathcal{E}], \forall t\in[1,T] \tag{19d}$$

The local feasibility cutting planes (19d) can be viewed as a further decoupling of the feasibility cutting plane (15e) at the MG level. The subscripts indicate that these parameters and variables belong to iteration $e$, time period $t$ and MG $n$. Therefore, we have $P_{e,t}^{fea} = P_{e,t,1}^{fea} \cup P_{e,t,2}^{fea} \cup \cdots \cup P_{e,t,\mathcal{N}}^{fea}$. After solving the LMP of each MG, the LB of the objective function of the whole distributed UC problem can be obtained by accumulating the LBs of the objective functions of the LMPs $\underline{Z}^{\varepsilon} = \sum_n \underline{Z}_n^{\varepsilon}$. Similar to QGBD, D-CQGBD converts NP-hard LMP into QUBO-LMP, which is handled by the quantum processor. QGBD cannot be expanded to the distributed version, because its feasible sub-problem does not contain the consensus information, making the generated feasibility Benders cutting planes (15e) undecouplable on MGs.

**Algorithm 2** outlines the procedures of D-CQGBD. The classical Gurobi9 solver is utilized to solve the local optimal economic planning sub-problems and relaxed feasible sub-problems within each MG. The LMP for each MG is processed through the quantum computing steps, which are highlighted in yellow. QGBD employs a centralized strategy, whereas D-CQGBD adopts a fully distributed solving strategy at the MG level. D-CQGBD decomposes the UC problem into local sub-problems and LMPs within each MG, and introduces the interaction information from neighboring MGs, represented as consensus variables. The consensus constraints ensure that after consensus iterations, all neighboring MGs reach a consensus, meaning that the residual of the consensus variables becomes sufficiently small to approximate the global optimum. These consensus steps are highlighted in green.

---

**Algorithm 2:** D-CQGBD

---

**Input:** Maximum iteration: $Iter_{max}$, cost coefficients: $a_i, b_i, g, d_i$, constraints coefficients: $A,B,C,D,E,F,G$,
number of MGs $\mathcal{N}$, the units contained in each MG $n\in[1,\mathcal{N}]$, error tolerance $\omega$;

1. $UB=+\infty, LB=-\infty$ , $\mathcal{E}=1$ , $U^{*}=random$ ; // Initialize.
2. **while** $UB-LB\geq\omega$ and $\mathcal{E}\leq Iter_{max}$ **do**:
3. $U$ in the distributed UC problem is fixed as $U^{*}$ to obtain the sub-problems;
4. **if** the sub-problem is feasible:
5. Introduce consensus variables into the adjacent MGs.
6. **While** each adjacent MG not reach a consensus **do**:
7. Solve the local sub-problem in parallel with $U_{n}=U_{n}^{*}$ and consensus constraints;
8. Obtain the local output power $P_{\mathcal{E},n}^{op}$ and consensus variables $\vec{P}_{n,m}^{n}$ of each MG;
9. Update global consensus variables $\vec{P}_{n,m}^{+}$ with average consensus strategy;
10. Calculate residuals and update dual variables of consensus constraints;
11. Obtain the local dual information of each MG $L_{\mathcal{E},n}^{op}, M_{\mathcal{E},n}^{op}, n\in[1,\mathcal{N}]$ ;
12. Obtain the local objective function $\bar{Z}_{n}^{\mathcal{E}}$ and set $\bar{Z}^{\mathcal{E}}=\sum_{n}\bar{Z}_{n}^{\mathcal{E}}$ ;
13. Set the continue variables solution $P_{\mathcal{E}}^{op}=P_{\mathcal{E},1}^{op}\cup\ldots\cup P_{\mathcal{E},\mathcal{N}}^{op}$ ;
14. **if** $\bar{Z}^{\mathcal{E}}\leq UB$ :
15. update $UB=\bar{Z}^{\mathcal{E}}$ ;
16. **end**
17. Construct multiple local optimality Benders cutting planes;
18. **else:**
19. Introduce relaxation variables $S$ and consensus constraints;
20. **While** each adjacent MG not reach a consensus **do**:
21. Construct and solve the local feasible sub-problems in parallel;
22. Obtain the local output power $P_{\mathcal{E},n}^{fea}$ , relaxation variables and consensus variables $\vec{P}_{n,m}^{n}$ ;
23. Update global consensus variables $\vec{P}_{n,m}^{+}$ with average consensus strategy;
24. Calculate residuals and update dual variables of consensus constraints;
25. Obtain the dual information of each MG $L_{\mathcal{E},n}^{fea}, M_{\mathcal{E},n}^{fea}, n\in[1,\mathcal{N}]$ ;
26. Construct multiple local feasibility Benders cutting planes;
27. **end**
28. Return multiple local optimality and feasibility Benders cutting planes to MP;
29. Decompose the MP into multiple LMPs corresponding to each MG $n\in[1,\mathcal{N}]$ ;
30. Transform LMP into QUBO-LMP by introducing auxiliary binary variables and penalties;
31. Transform QUBO-LMP into the problem Hamiltonian $H_{P}$ ;
32. Map the Ising model to quantum annealing devices;
33. Solve QUBO-LMP with quantum annealing by the QPU in each MG;
34. Obtain the local solution $U_{\mathcal{E},n}$ for each MG;
35. Set the binary variables solution $U_{\mathcal{E}}=U_{\mathcal{E},1}\cup\ldots\cup U_{\mathcal{E},\mathcal{N}}$ ;
36. Set $\underline{Z}^{\mathcal{E}}=\sum_{n}\underline{Z}_{n}^{\mathcal{E}}=\sum_{n}min\left\{c_{e,n}^{op}+\left(f_{e,n}^{op}\right)^{T}\cdot U_{\mathcal{E},n}\middle|e=1,\ldots,\mathcal{E}\right\}$ ;
37. **if** $\underline{Z}^{\mathcal{E}}\geq LB$ :
38. update $LB=\underline{Z}^{\mathcal{E}}$ ;
39. **end**
40. Update $U^{*}=U_{\mathcal{E}}$ ;
41. $\mathcal{E}=\mathcal{E}+1$ ;
42. **end while**;

**Output:** Hourly on/off states $U=U^{*}$ , output power $P=P_{\mathcal{E}}^{op}$ of the units, the operation charges $UB$ .

### 4.3. Framework of D-CIQGBD

In this subsection, we decompose the QUBO-MP into multiple QUBO-LMPs based on the inspired consensus among units, and then develop a partially distributed version D-CIQGBD. **Figure 4** depicts the architecture of D-CIQGBD. The whole system consists of several subsystems equipped with local quantum processors. The LMP of each MG is solved by its own quantum processor, and the unit on/off status is obtained via quantum measurement, leading to a parallel and distributed computation. The economic dispatching coordinator uses the classical optimizer to solve the sub-problems in a centralized manner, and obtains the optimal output powers or relaxation powers of units in each MG under the current unit on/off status, which are returned to MGs in the form of cutting planes and result in new LMPs.

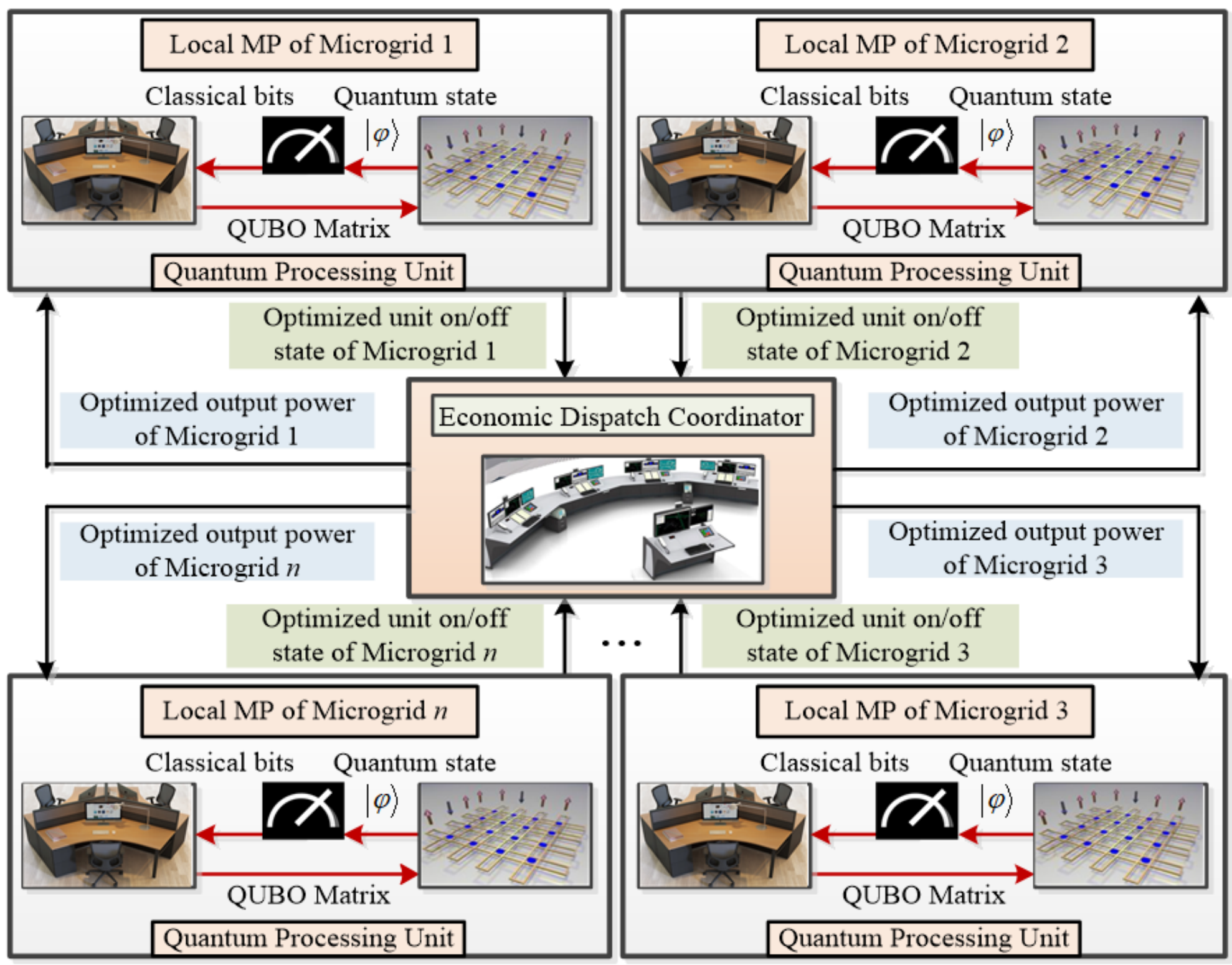

**Figure 4.** Partially distributed coordination architecture of D-CIQGBD.

---

**Algorithm 3:** D-CIQGBD

---

**Input:** Maximum iteration: $Iter_{max}$ , cost coefficients: $a_i,b_i,g,d_i$ , constraints coefficients: $A,B,C,D,E,F,G$ ,
number of MGs $\mathcal{N}$ , the units contained in the MG $n\in[1,\mathcal{N}]$ , error tolerance $\omega$ ;

1. $UB=+\infty, LB=-\infty$ , $\mathcal{E}=1$ , $U^*=random$ ; // Initialize.
2. **while** $UB-LB\geq\omega$ and $\mathcal{E}\leq Iter_{max}$ **do**:
3. $U$ in the distributed UC problem is fixed as $U^*$ to obtain the sub-problems;
4. **if** the sub-problem is feasible:
5. Solve the sub-problem with $U=U^*$ ;
6. Obtain the output power $P_{\mathcal{E}}^{op}$ ;
7. Obtain the local output power $P_{\mathcal{E},n}^{op}$ according to $P_{\mathcal{E}}^{op}=P_{\mathcal{E},1}^{op}\cup\ldots\cup P_{\mathcal{E},\mathcal{N}}^{op}$ ;
8. Obtain the local dual information $L_{\mathcal{E},n}^{op},M_{\mathcal{E},n}^{op},n\in[1,\mathcal{N}]$ ;
9. Obtain the objective function $\bar{Z}^{\mathcal{E}}$ ;
10. **if** $\bar{Z}^{\mathcal{E}}\leq UB$ :
11. update $UB=\bar{Z}^{\mathcal{E}}$ ;
12. **end**
13. Construct multiple local optimality Benders cutting planes;
14. **else:**
15. Introduce relaxation variables $S$ , consensus-inspired cost function and constraints;
16. Construct and solve the consensus-inspired feasible sub-problems;
17. Obtain the continue variables solution of each MG $P_{\mathcal{E},n}^{fea},n\in[1,\mathcal{N}]$ ;
18. Obtain the dual information of each MG $L_{\mathcal{E},n}^{fea},M_{\mathcal{E},n}^{fea},n\in[1,\mathcal{N}]$ ;
19. Construct multiple local feasibility Benders cutting planes;
20. **end**
21. Return multiple local optimality and feasibility Benders cutting planes to MP;
22. Decompose the MP into multiple LMPs corresponding to each MG $n\in[1,\mathcal{N}]$ ;
23. Transform LMP into QUBO-LMP by introducing auxiliary binary variables and penalties;
24. Transform QUBO-LMP into the problem Hamiltonian $H_P$ ;
25. Map the Ising model to quantum annealing devices;
26. Solve QUBO-LMP with quantum annealing by the QPU in each MG;
27. Obtain the local solution $U_{\mathcal{E},n}$ for each MG;

28. Set the binary variables solution $U_{\mathcal{E}}=U_{\mathcal{E},1}\cup\ldots\cup U_{\mathcal{E},\mathcal{N}}$;
29. Set $\underline{Z}^{\mathcal{E}}=\sum_n \underline{Z}_n^{\mathcal{E}}=\sum_n min\left\{c_{e,n}^{op}+\left(f_{e,n}^{op}\right)^T\cdot U_{\mathcal{E},n} \middle| e=1,\ldots,\mathcal{E}\right\}$;
30. **if** $\underline{Z}^{\mathcal{E}}\geq LB$:
31. update $LB=\underline{Z}^{\mathcal{E}}$;
32. **end**
33. Update $U^*=U_{\mathcal{E}}$;
34. $\mathcal{E}=\mathcal{E}+1$;
35. **end while**;

**Output:** Hourly on/off states $U=U^*$, output power $P=P_{\mathcal{E}}^{op}$ of the units, the operation charges $UB$.

In QGBD, the optimality cutting planes (15c), obtained by solving the optimal sub-problem (13), already incorporate the consensus information (*i.e.*, $\mathcal{K}(P)$ in (13a)), making (15c) decouplable on MGs. However, the feasible sub-problem (14) only considers the global information of the entire power system. As a result, it lacks the capability to perform decoupling operations or handle specific interactions among MGs. This limitation makes QGBD less suitable for complex power systems with distributed energy resources, where local decision-making and decentralized optimization are crucial. Inspired by the consensus among the units, D-CIQGBD improves QGBD via giving cost weights to the relaxation variables $S$ of the feasible sub-problem (14), leading to (21b). Given the Lagrange function of the feasible sub-problem as:

$$\begin{aligned}\mathcal{L}\left(S,P,L,M,K\right)=&\|{}^1S\|_1+\|{}^2S\|_1+\left(L^{fea}\right)^T\left(C\cdot P-D-{}^1S\right)\\&+\left(M^{fea}\right)^T\left(E\cdot U^*+F\cdot P-G-{}^2S\right)-\left({}^1K^{fea}\right)^T{}^1S-\left({}^2K^{fea}\right)^T{}^2S,\end{aligned}\tag{20}$$

the following consensus-inspired feasible sub-problem can be constructed:

$$\min\sum_{i=1}^{I}\sum_{t=1}^{T}\mathcal{K}_{i,t}\left(S\right),\tag{21a}$$

$$\mathcal{K}_{i,t}\left(S\right)=a_i\cdot {s_{i,t}}^2+b_i\cdot s_{i,t}+g_i,\tag{21b}$$

$$s.t.\quad KKT\ condition\ of\ the\ feasible\ sub-problem\ (5)\tag{21c}$$

$$S\leq P^{\max}\left(V-U^*\right).\tag{21d}$$

Here $S\in\left\{{}^1S,{}^2S\right\},s_{i,t}\in\left\{{}^1s_{i,t},{}^2s_{i,t}\right\}$, and ${}^1S,{}^2S\in\boldsymbol{R}^{IT\times 1}$ are the column vectors with elements of relaxation variables ${}^1s_{i,t},{}^2s_{i,t}$, respectively. $V$ is a column vector of size $IT$ with all elements being 1. Deviations ${}^1s_{i,t},{}^2s_{i,t}$, which are essentially the relaxation powers of unit *i*, are respectively added to (14b) and (14c) to make unit *i* satisfy constraints (14b) and (14c) on the time period *t*.

The function of (21a) is to make the units with lower operation costs get more relaxation, which means greater tolerance for violating constraints. The purpose of (21d) is to relax the "off" units, while not relaxing the "on" units, and the relaxation power satisfies the maximum power constraints. The feasible sub-problem (21) aims to make the first derivative of the cost function of relaxation variables $\nabla_{s_{i,t}}\mathcal{K}_{i,t}\left(S\right)=\left(2a_i\cdot s_{i,t}+b_i\right)$ for each unit on the same time period *t* as consistent as possible, which means that units with lower operation costs undertake more power generation tasks. By solving (21), the dual information ($L^{fea}$, $M^{fea}$) and feasible solution of continuous variables ($P^{fea}$) in each MG can be obtained. The feasibility Benders cutting planes (15e) can be further decoupled on different MGs, leading to (19d).

Therefore, D-CIQGBD decomposes the MP into LMPs of different MGs, effectively decoupling the constraints of binary variables, optimality and feasibility cutting planes. The procedures of D-CIQGBD are described in **Algorithm 3**. The difference between D-CIQGBD and CIQGBD lies in their strategies for solving the MP. In D-CIQGBD, the MP with complex binary variables can be efficiently decoupled into LMPs within each MG and then solved in parallel. This decoupling process is similar to the LMP processing step in D-CQGBD (*i.e.*, (19)). On the other hand, CIQGBD, the centralized version of D-CIQGBD, adopts a centralized approach to solve the MP. In CIQGBD, the MP involves summing up the objective functions and combining the constraints from all LMPs. Compared to QGBD, CIQGBD replaces the original $\mathcal{E} \times T$ feasibility Benders cutting planes with $\mathcal{E} \times T \times \mathcal{N}$ local feasibility Benders cutting planes of all MGs, and reduces the constraint complexity by increasing the number of cutting planes, thus improving the QUBO-MP.

### 4.4. Ising Reformulation of the MP/LMP

The fundamental units of quantum devices are qubits. One qubit can be in the superposition of states $|0\rangle$ and $|1\rangle$. After measurement, qubits will collapse into $|0\rangle$ or $|1\rangle$ with different probabilities. A quantum system composed of multiple qubits can also have entanglement. For example, an entangled quantum system consisting of two qubits is in the superposition of four entangled states (*i.e.,* $|0\rangle|0\rangle, |0\rangle|1\rangle, |1\rangle|0\rangle$ and $|1\rangle|1\rangle$). Therefore, a system consisting of $n$ qubits will have $2^n$ possible states. These quantum characteristics may bring some special advantages in reducing computational complexity.

The processes of reconstructing MP of QGBD and LMP of D-CQGBD/D-CIQGBD to the QUBO form are similar, so this subsection takes the case of reconstructing MP as an example. In the following, we first discuss the detailed process of converting the MP into the QUBO-MP, and then describe how to map the QUBO-MP to the QAOA quantum circuit or quantum annealing machine.

The optimality and feasibility Benders cutting planes in (15c) and (15e) can be rewritten in the following form:

$$Z^{\mathcal{E}} \geq c_e^{op} + \left(f_e^{op}\right)^T \cdot U, \forall e \in [1, \varepsilon] \tag{22a}$$

$$0 \geq c_{e,t}^{fea} + \left(f_{e,t}^{fea}\right)^T \cdot U_t, \forall e \in [1, \varepsilon], \forall t \in [1, T] \tag{22b}$$

Non-negative auxiliary integer variables $S^{cons} \in \boldsymbol{N}^{O \times I}$ and $S_{e,t}^{fea} \in \boldsymbol{N}$ expressed in binary representation are introduced to transform the above inequality constraints into equality binary constraints:

$$A \cdot U - B + S^{cons} = 0, \tag{23a}$$

$$c_{e,t}^{fea} + \left(f_{e,t}^{fea}\right)^T \cdot U_t + S_{e,t}^{fea} = 0, \ \forall e \in [1, \varepsilon], \forall t \in [1, T] \tag{23b}$$

Then the constraints (23a) and (23b) are, respectively, added as penalty terms (24d) and (24c) in the objective function of the QUBO-MP, thus ensuring that the constraints are satisfied.

By making $\underline{Z}^{\varepsilon}$ (*i.e.,* the LB of the initial problem) gradually converge to $\overline{Z}^{\varepsilon}$ (*i.e.,* the UB of the initial problem), the objective function of the MP can be rewritten in the QUBO form (24b), where $\mu$ and $\eta$ are adjustable parameters [16]. Therefore, the original MP is reconstructed into the following QUBO-MP:

$$\min H = H_{op} + H_{fea} + H_{cons} \tag{24a}$$

$$s.t. \quad H_{op} = \left[\sum_e \left(f_e^{op} + 2\mu\left(c_e^{op} - \overline{Z}_{\varepsilon}\right) f_e^{op}\right)\right]^T U + U^T \left[\eta \sum_{e,t,t'} f_{e,t}^{op} \left(f_{e,t'}^{op}\right)^T\right]_{[i,j]=0, i \geq j} U, \tag{24b}$$

$$H_{fea} = \xi^{fea} \sum_{e \in \mathcal{E}} \sum_{t \in T} \left( c_{e,t}^{fea} + \left(f_{e,t}^{fea}\right)^T \cdot U + S_{e,t}^{fea} \right)^2 \tag{24c}$$

$$H_{cons} = \xi^{cons} \left\| A \cdot U - B + S^{cons} \right\|_2^2. \tag{24d}$$

Here $[A]_{[i,j]=0, i \geq j}$ is obtained by making the lower triangle matrix elements of $A$ be zero. The objective function in the QUBO form $H$ takes binary variables $U$ as decision variables, and consists of three different terms. $H_{op}$ counts for the objective function of the MP (15a), while $H_{fea}$ and $H_{cons}$ are for constraints.

QGBD converts the information into the coefficients of linear and quadratic terms of the QUBO-MP, and the numerical values of diagonal and non-diagonal elements (corresponding to linear and quadratic term coefficients, respectively) in the QUBO matrix change dynamically with iterations. Therefore, the size of the QUBO matrix does not increase linearly with iterations, and increases when only some auxiliary binary variables have to be introduced from the added feasibility cutting planes. The QUBO-MP (24) has the following form:

$$H = \sum_{(i,j)} \rho_{i,j} U_i U_j + \sum_i v_i \, U_i + a, \tag{25}$$

which can be readily converted into the Ising model of quantum computing.

The Ising model can be defined as a graph $G(L,K)$ with vertices set $L$ and edges set $K$. The vertices can be regarded as qubits in the system and the edges denote possible interactions between qubits. The Hamiltonian of the Ising model is：

$$H_p = \sum_{(i,j)} k_{i,j} \hat{\sigma}_z^{(i)} \hat{\sigma}_z^{(j)} + \sum_i h_i \, \hat{\sigma}_z^{(i)}. \tag{26}$$

Here $\hat{\sigma}_z = \begin{bmatrix} 1 & 0 \\ 0 & -1 \end{bmatrix}$ is the Pauli-Z operator, and the superscript $i$ indicates that this operator acts on qubit $i$. The first term of $H_p$ counts for the interaction between qubits, and the second term of $H_p$ are for the interaction between the external magnetic field and the qubit.

The objective function *H* in Eq. (25) can be transformed to the Ising Hamiltonian (26) via the following mapping between variables $U \in \{0,1\}^{IT \times 1}$ and Pauli-Z operators in the Ising model:

$$U_i \Leftrightarrow \frac{1}{2}(I - \hat{\sigma}_z^{(i)}), \forall i \in [1, IT]. \tag{27}$$

When the spin of qubit $i$ is measured to be 1 (*i.e.*, $|i\rangle = |0\rangle = \begin{pmatrix} 1 \\ 0 \end{pmatrix}$), that is, $\langle i | \hat{\sigma}_z^{(i)} | i \rangle = 1$, the expectation value of the operator on the right hand side of Eq. (27) is $\left\langle i | \frac{1}{2}(I - \hat{\sigma}_z^{(i)}) | i \right\rangle = 0$, which is just taken as $U_i = 0$ (*i.e.*, unit "off"); When the spin of qubit $i$ is measured to be -1 (*i.e.*, $|i\rangle = |1\rangle = \begin{pmatrix} 0 \\ 1 \end{pmatrix}$), the expectation value of the operator is $\left\langle i | \frac{1}{2}(I - \hat{\sigma}_z^{(i)}) | i \right\rangle = 1$, which is just taken as $U_i = 1$ (*i.e.*, unit "on").

The Ising Hamiltonian transformed from the QUBO-MP can be implemented with the QAOA quantum circuit or the quantum annealing machine. Then QAOA or quantum annealing can be used to obtain the ground state of the Ising Hamiltonian, which has the lowest energy corresponding to the minimal operation cost of the power system. At the same time, the expectation value of $\frac{1}{2}(I - \hat{\sigma}_z^{(i)})$ on each qubit just gives the unit on/off solution of the MP.

## 5. Numerical Results

This section illustrates the performance of QGBD, CIQGBD, D-CQGBD and D-CIQGBD in solving different UC problems. In these algorithms, the classical solver Gurobi9 is used to solve the sub-problem, while the QAOA algorithm or quantum annealing algorithm is used to solve the QUBO-MP and QUBO-LMP. The QAOA and quantum annealing algorithms are performed on the IBM Qiskit and D-WAVE $Leap^{TM}$ platforms, respectively. In the following experiments, the solution time of the decomposition algorithms consists of the time taken to solve the MP and the subproblems. Specifically, the solution time for solving the QUBO-MP via quantum annealing on D-WAVE includes both the QPU programming time and the QPU sampling time.

Since the maximal number of available qubits in IBM Qiskit is limited, it cannot satisfy the requirements of addressing the QUBO-MP with QAOA when considering the generator minimum on/off time constraints or a large-scale MG. Therefore, in the centralized scenarios, when QAOA is used as an additional option beyond quantum annealing in QADMM and QGBD, we take an MG with three DERs and no generator minimum on/off time constraints as an example (*i.e.*, scenario 1).In the distributed scenarios, where the scale of MG is larger and generator minimum on/off time constraints are considered in addition to system power demand constraints and generator output power constraints, only the quantum annealing algorithm is used to tackle the QUBO problem. The time periods for all the following scenarios are 24 hours. Finally, we further incorporate system power flow constraints and the transmission line power flow capacity constraints into the UC problem. This more realistic SCUC model significantly increases the computational complexity and, consequently, the solution time. We use this more complex SCUC model to test the performance advantage of D-CIQGBD relative to Gurobi9.

### 5.1. Centralized Scenarios

Cooperation between multiple networked MGs and the local utility grid (LUG) exists in the centralized UC problem, and the strategy of global modeling is adopted to solve the UC problem.

#### *5.1.1. Scenario 1*

An MG consisting of three DERs is considered. In this scenario, the UC problem considers only the system power balance constraint, and minimum and maximum unit output power constraints. When quantum annealing is used to solve the QUBO problem, unit minimum on/off time constraints are also considered. Parameters of each DER and the hourly power demand profile are given in ref.[31].

When the error limits of GBD, QGBD, and the convergence residual of QADMM are all set to $10^{-4}$ , these algorithms can converge to the same result (shown in **Table 1**) as Gurobi9: DER1 and DER3 are in on status from the very beginning and participate in the whole power generation process; DER2, which is the least economical, participates in the power generation process from the 7-th to the 22-th hour. Because generator minimum on/off time constraints are not considered in scenario 1, the UC problem is decouplable on $t$. When QADMM, GBD, and QGBD solve the UC problem in different time periods, the numbers of iterations to achieve convergence are shown in **Figure 5**. It can be seen that the number of iterations required by GBD and QGBD is the same and the least, while QADMM requires more iterations.

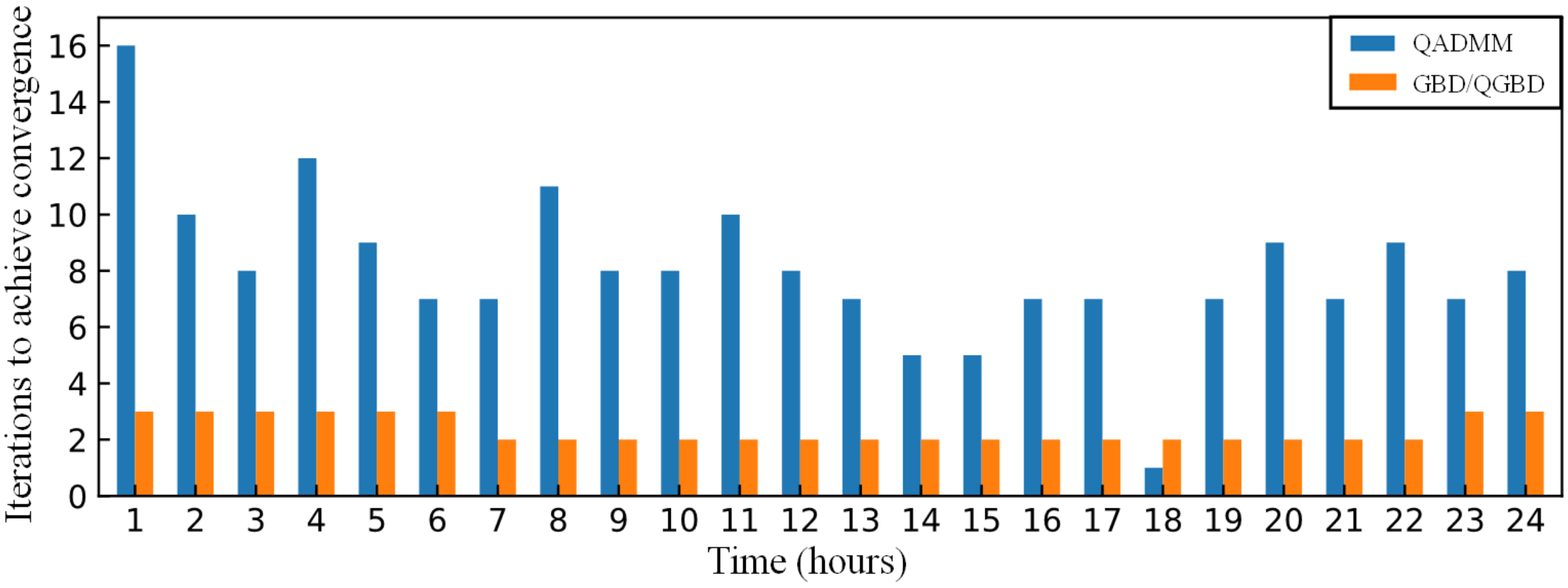

**Figure 5.** Iterations of QADMM, GBD and QGBD to achieve convergence for different time periods.

**Table 1.** Result of UC problem in the first scenario (kW)

| DER No. | T1 | T2 | T3 | T4 | T5 | T6 | T7 | T8 | T9 | T10 | T11 | T12 |
|---|---|---|---|---|---|---|---|---|---|---|---|---|
| 1 | 15 | 15 | 15 | 15 | 15 | 15 | 15 | 15 | 15 | 15 | 15 | 15 |
| 2 | 0 | 0 | 0 | 0 | 0 | 0 | 4 | 7 | 12 | 9 | 6 | 5 |
| 3 | 1 | 3 | 5 | 7 | 10 | 15 | 15 | 15 | 15 | 15 | 15 | 15 |
| Demand | 16 | 18 | 20 | 22 | 25 | 30 | 34 | 37 | 42 | 39 | 36 | 35 |
| DER No. | T13 | T14 | T15 | T16 | T17 | T18 | T19 | T20 | T21 | T22 | T23 | T24 |
| 1 | 15 | 15 | 15 | 15 | 15 | 15 | 15 | 15 | 15 | 15 | 15 | 15 |
| 2 | 4 | 11 | 11 | 10 | 13 | 15 | 10 | 8 | 4 | 2 | 0 | 0 |
| 3 | 15 | 15 | 15 | 15 | 15 | 15 | 15 | 15 | 15 | 15 | 15 | 9 |
| Demand | 34 | 41 | 41 | 40 | 43 | 45 | 40 | 38 | 34 | 32 | 30 | 24 |

### 5.2. Distributed Scenarios

In the following four distributed scenarios, where the generator minimum on/off time constraints are considered in addition to system power demand constraints and generator output power constraints, only the quantum annealing algorithm is used to tackle the QUBO problem. The time periods for all the following scenarios are 24 hours. Consensus-inspired distributed methods D-CIGBD and D-CIQGBD as well as their centralized versions CIGBD and CIQGBD are applied to the UC problem in scenarios 2-5. Parameters of the DERs and the hourly power demand profile are given in ref. [31].

#### *5.2.1. Scenario 2*

We examine a scenario involving 6 DERs distributed across 3 MGs to discuss the compactness of the local cutting planes in D-CGBD and D-CIGBD at the numerical level. We aim to illustrate the differences in cutting planes and the convergence of these two algorithms using a three-dimensional visualization. To simplify the analysis, we focus on a single time period of the distributed UC problem, excluding unit minimum on/off time constraints. Specifically， in each MG, two DERs are positioned, and the on/off status of these DERs serves as the decision variables for the LMP. The system satisfies the power balance constraints, as well as the minimum and maximum unit output power constraints.

**Figure 6** presents a three-dimensional plot, illustrating the cutting plane data for the LMPs within each of the three MGs at each iteration. The subplots in each row represent the cutting planes progressively added to the LMPs of MG1, MG2, and MG3 as the iterations proceed. The $\mathcal{X},\mathcal{Y}$ axes in the plot correspond to the binary variables of the two DERs.

The right-hand sides of the local feasibility cutting plane (19d) and local optimality cutting plane (19c) are denoted as $\underline{\mathcal{L}}\left(U_n;L_n^{fea};M_n^{fea};P_n^{fea}\right)$ and $\bar{\mathcal{L}}\left(U_n;L_n^{op};M_n^{op};P_n^{op}\right)$, respectively. The binary variable $U_n$ serves as the decision variable, while the Lagrange multipliers and the continuous variable solution act as hyperparameters. In **Figure 6**, the black plane represents the plane $\mathcal{Z}=0$. The red and blue planes represent the local feasibility cutting planes $\mathcal{Z}=\underline{\mathcal{L}}\left(U_n;L_n^{fea};M_n^{fea};P_n^{fea}\right)$ added by D-CIGBD and D-CGBD, respectively. The yellow and green planes depict the local optimality cutting planes $\mathcal{Z}=\bar{\mathcal{L}}\left(U_n;L_n^{op};M_n^{op};P_n^{op}\right)$ during the second and third iterations, respectively. Based on the constraints $0\geq\underline{\mathcal{L}}\left(U_n;L_n^{fea};M_n^{fea};P_n^{fea}\right)$, the feasible region of the LMP is the part below the black plane that intersects with the local feasibility cutting planes. The red and blue dots respectively indicate the feasible regions of D-CIGBD and D-CGBD. The feasible region of D-CGBD is smaller than that of D-CIGBD, indicating that the local feasibility cutting planes constructed by D-CGBD are more compact than those of D-CIGBD. Based on the feasible region and the local optimal constraints $Z_n \geq \bar{\mathcal{L}}\left(U_n; L_n^{op}; M_n^{op}; P_n^{op}\right)$, the point with the minimum objective function value $\underline{Z}_n = \min_{U_n} Z_n$ is marked with orange asterisks.

Importantly, whether the feasible region is obtained using D-CGBD or D-CIGBD, both methods can attain the same optimal solution by considering the local optimality cutting planes. We reach the conclusion that D-CGBD constructs more compact local feasibility cutting planes. However, both algorithms generate the same local optimality cutting planes. For UC problems of the same scale, the convergence process of D-CIQGBD and D-CGBD is theoretically identical, as shown in Figure 6. However, D-CIQGBD allocates slack variables more reasonably based on unit cost information, eliminating the need for the time-consuming average consensus iterative process in D-CQGBD. As a result, the D-CIQGBD subproblems reach consensus more quickly. This advantage is also validated in Scenario 4.

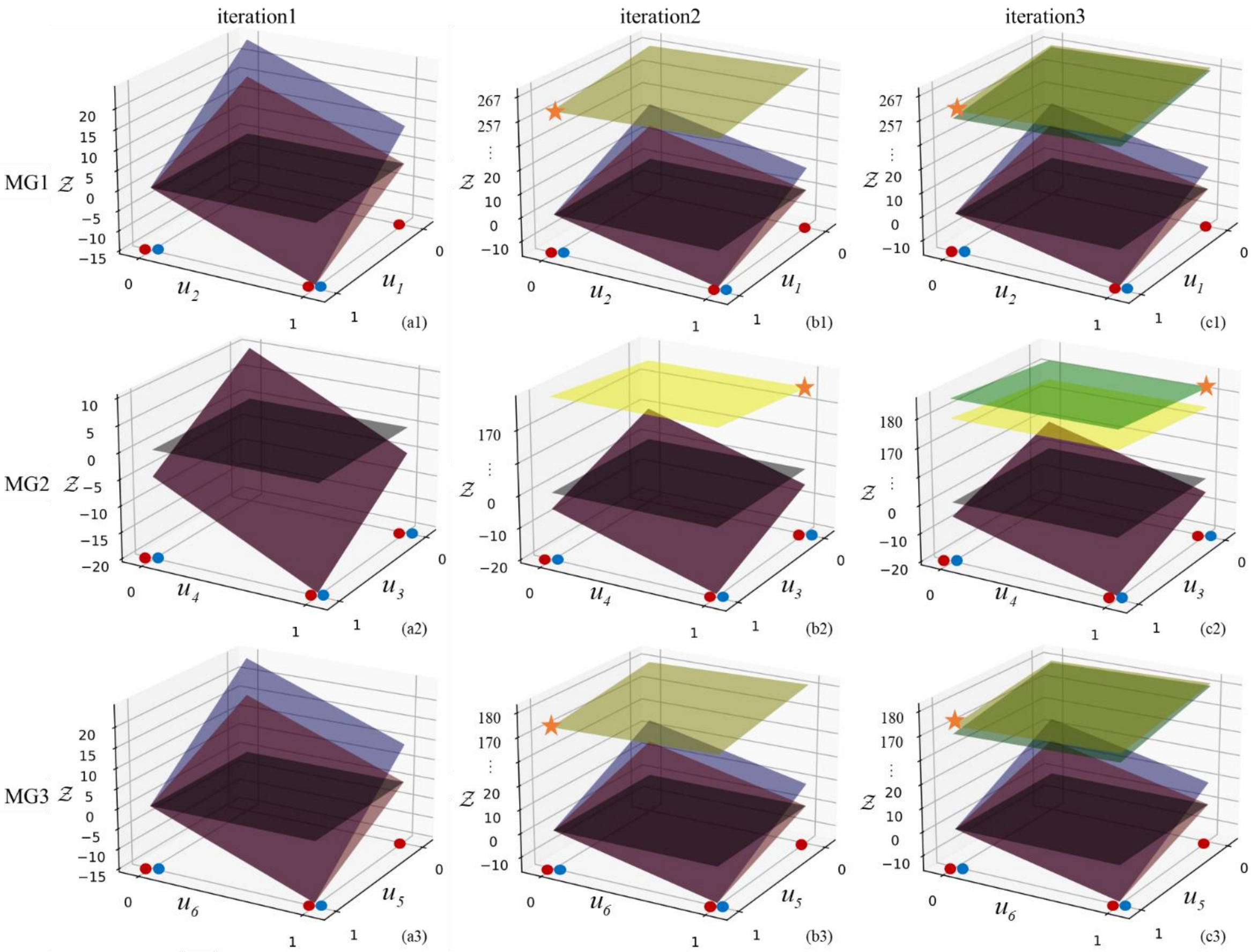

**Figure 6.** Three-dimensional plot of the cutting planes for the LMP in the UC problem with 6-DERs and 3-MGs. The red and blue dots respectively represent the feasible regions of D-CIGBD and D-CGBD. The red and blue planes illustrate the feasibility cutting planes. The yellow and green planes depict the optimality cutting planes, and the orange asterisk symbolizes the optimal solution.

*5.2.2. Scenario 3*

The power is generated by 9 schedulable DERs cooperatively. Parameters of each DER and the hourly power demand profile are given in ref.[31]. In this scenario, the UC problem only considers the physical constraints of the unit: system power balance constraint, minimum and maximum unit output power constraints, and unit minimum on/off time constraints. The results are listed in **Table 2**. In the centralized methods, Gurobi9 takes the strategy of global modeling for solving, while GBD, CIGBD, and their quantum versions QGBD and CIQGBD solve the MP and sub-problems in a coordinated way. In the distributed methods, both D-CGBD and its quantum version D-CQGBD fully distribute the local sub-problems and LMPs of individual

MGs for parallel solving. On the other hand, D-CIGBD and its quantum version D-CIQGBD solve the LMPs of different MGs in a distributed and parallel way. It should be noted that due to the fact that the solution of different LMPs is independent in principle, LMPs can be calculated in parallel at the same time.. This is different from the distributed version of QADMM, where the QUBO problem of one MG takes the solution of QUBO problems of the other MGs as input parameters. In this way, different QUBO problems must be solved one by one in principle.

As shown in **Table 2**, the operational costs obtained by GBD, QGBD, and consensus-inspired algorithms (CIGBD/CIQGBD/D-CIGBD/D-CIQGBD) are the same and lower than that of Gurobi9, and slightly lower than the costs obtained by consensus algorithms (D-CGBD/D-CQGBD). This outcome can be attributed to the fact that consensus algorithms converge based on the consensus agreement among neighboring MGs, where the convergence criterion is set as the sufficiently small residual between the local consensus variables and the global variables. In these examples, we adopt a convergence criterion of a residual of $10^{-4}$, which could introduce errors and thereby impact the final results, leading to a tendency to converge towards approximate optimal values.

**Figure 7** (a) and (b) give the on/off status and the output powers of each DER within 24 hours obtained by these algorithms. The convergence behaviors of GBD, QGBD, CIGBD, CIQGBD, D-CGBD, D-CQGBD, D-CIGBD and D-CIQGBD are compared in **Figure 7** (c) and (d). CIGBD, CIQGBD, D-CIGBD, D-CIQGBD, D-CGBD, and D-CQGBD converge in 3 iterations to obtain a high-quality feasible solution, while QGBD and GBD converge in 5 iterations to obtain the same results.

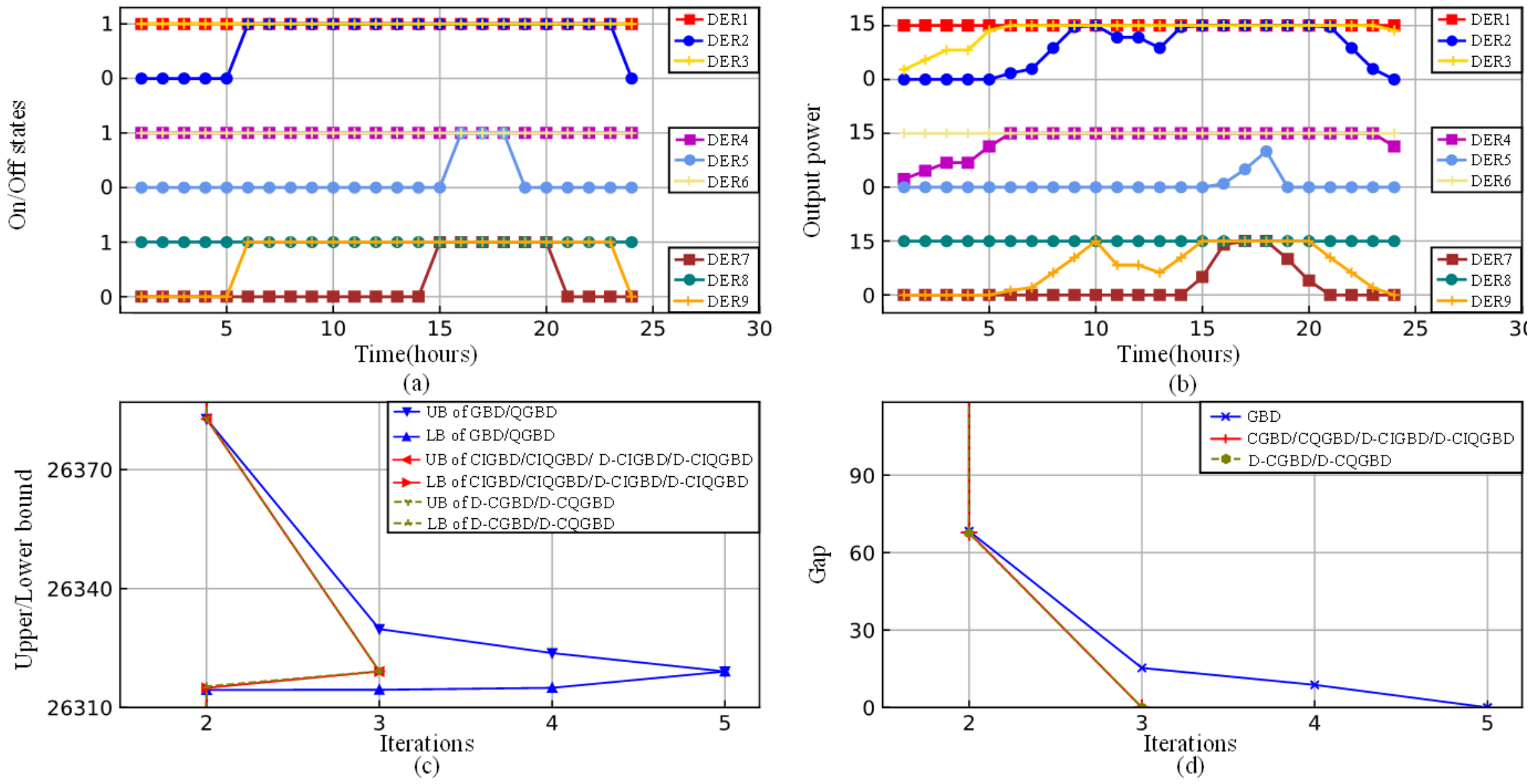

**Figure 7.** (a)24-hour unit on/off status, (b) 24-hour output power, (c) the UBs and LBs, and (d) the gap between UB and LB in Scenario 3.

**Table 2.** Operation costs in Scenario 3

| MG No. | Gurobi9 | GDB | QGBD | CIGBD/CIQGBD/D-CIGBD/D-CIQGBD | D-CGBD/D-CQGBD |
|---|---|---|---|---|---|
| Number of iterations | —— | 5 | 5 | 3 | 3 |
| MG1 | \$10,559.57 | \$10,568.38 | \$10,568.38 | \$10,568.38 | \$10,545.92 |
| MG2 | \$7,518.21 | \$7,518.21 | \$7,518.21 | \$7,518.21 | \$7,518.84 |
| MG3 | \$8,241.45 | \$8,232.49 | \$8,232.49 | \$8,232.49 | \$8,253.39 |
| Total | \$26,319.23 | \$26,319.08 | \$26,319.08 | \$26,319.08 | \$26,319.15 |

*5.2.3. Scenario 4*

To investigate the performance improvement brought by the consensus-inspired distributed strategy, we increase the number of DERs from 18 to 63 in this scenario, and use centralized and distributed classical/quantum algorithms to solve these UC problems at different scales. In this scenario, the UC problem only considers the physical constraints of the unit.

The number of iterations, operation costs, and solution time by Gurobi9, GBD, CIGBD, D-CIGBD, D-CGBD, CIQGBD, D-CIQGBD, and D-CQGBD are compared in **Table 3**. **Figure 8** shows the convergence behaviors of the UBs and LBs of the UC problems at four scales. In principle, partially distributed algorithms (D-CIGBD/D-CIQGBD) can handle the LMPs in parallel, while fully distributed algorithms (D-CGBD/D-CQGBD) can parallelly handle both the local sub-problems and LMPs. However, due to hardware constraints, we have opted for a sequential approach, solving the LMPs and local sub-problems one by one according to the order of MGs. To estimate the theoretical solving time of these distributed algorithms, we divide the time taken for sequential solving by the number of MGs that could be solved in parallel. This approach allows us to approximate the potential solving time improvements these distributed algorithms could achieve through parallelism. When the number of DERs increases from 27 to 63, CIGBD, CIQGBD, D-CIGBD, D-CIQGBD, D-CGBD, and D-CQGBD can all converge to high-quality solutions, and their operation costs are equal to or slightly less than that obtained by Gurobi9.

In centralized methods, the number of iterations of CIGBD and CIQGBD is obviously less than that of GBD, and CIQGBD is faster than CIGBD. Under the distributed framework, D-CIGBD and its quantum version D-CIQGBD, as well as D-CGBD and its quantum version D-CQGBD, require the least number of iterations. D-CIQGBD is faster than D-CIGBD for systems with more than 45 DERs. In the consensus algorithms D-CGBD and D-CQGBD, as the size of the UC problem and the number of MGs increase, the number of consensus variables also grows. This can lead to a higher likelihood of errors, as the local consensus variables within each neighboring MG may achieve consensus with some errors, contributing to an overall increase in errors. Additionally, achieving consensus coordination becomes more challenging with the enlargement of the size of units. To tackle this issue, designing more advanced consensus variable update strategies or modifying the MG topological structures is essential. Such improvements can enhance the accuracy and reliability of the consensus algorithms, even in scenarios with larger UC problems and more MGs. Compared to the average consensus algorithms D-CGBD and D-CQGBD, the subproblem solution time of the consensus-inspired algorithms D-CIGBD and D-CIQGBD is significantly reduced. This improvement is attributed to the consensus-inspired strategy, which employs an enhanced objective function of the feasible subproblem that incorporates unit cost information, enabling a more rational allocation of relaxation variables. This strategy allows feasible subproblems to obtain the optimal relaxation variable solutions under global optimization in a single step. In contrast, average consensus algorithms require multiple iterative adjustments of local consensus variables among subproblems to eventually align them with the global consensus variables. Although this iterative process can be executed in parallel, the computational burden of repeated adjustments remains substantial.

The results of QGBD are not listed in **Table 3**, because they are incorrect for systems with more than 18 DERs. The correctness of quantum annealing depends on the minimum gap between the ground state and the first excited state of the system in the annealing process, which is closely related to the rationality of the constructed problem Hamiltonian. Taking the UC problem of three DERs and a one-hour time period as an example, we use the simulated quantum annealing method to solve the QUBO-MP in CIQGBD and QGBD respectively, and obtain the eigenenergy levels and thus the eigenenergy gap between the ground state and the first excited state of the mapped Hamiltonians during the whole annealing process as shown in **Figure 9**, where the abscissa is the normalized anneal fraction in the range from 0 to 1.

**Table 3.** Performance comparison of different algorithms

| System | Data | Classical | | | | | Quantum | | |
|---|---|---|---|---|---|---|---|---|---|
| | | Gurobi9 | GBD | CIGBD | D-CIGBD | D-CGBD | CIQGBD | D-CIQGBD | D-CQGBD |
| 18-24 | Objective function value | 52638.17 | 52638.17 | 52638.17 | 52638.17 | 52638.43 | 52638.17 | 52638.17 | 52638.43 |
| | Number of iterations | —— | 5 | 3 | 3 | 3 | 3 | 3 | 3 |
| | Solution time (s) | 0.16 | 6.42 | 204.21 | 7.24 | 198.46 | 25.47 | 7 | 198.48 |
| | Solution time of MP (s) | —— | 4.67 | 203.04 | 6.05 | 6.05 | 24.46 | 7 | 5.98 |
| | Solution time of sub-problerm (s) | 0.16 | 1.75 | 1.17 | 1.19 | 192.41 | 1.01 | 1.01 | 192.50 |
| | Number of MGs | —— | —— | —— | 6 | 6 | —— | 6 | 6 |
| 27-24 | Objective function value | 78957.33 | 78957.25 | 78957.25 | 78957.25 | 78957.85 | 78957.25 | 78957.25 | 78957.85 |
| | Number of iterations | —— | 6 | 3 | 3 | 3 | 3 | 3 | 3 |
| | Solution time (s) | 0.21 | 19.42 | 320.18 | 8.64 | 198.20 | 29.97 | 7.77 | 197.34 |
| | Solution time of MP (s) | —— | 16.18 | 318.33 | 6.64 | 6.63 | 28.18 | 5.98 | 5.98 |
| | Solution time of sub-problerm (s) | 0.21 | 3.24 | 1.85 | 2 | 191.57 | 1.79 | 1.79 | 191.36 |
| | Number of MGs | —— | —— | —— | 9 | 9 | —— | 9 | 9 |
| 36-24 | Objective function value | 105276.40 | 105276.33 | 105276.33 | 105276.33 | 105276.35 | 105276.33 | 105276.33 | 105276.35 |
| | Number of iterations | —— | 7 | 4 | 3 | 3 | 4 | 3 | 3 |
| | Solution time (s) | 0.28 | 56.58 | 1121.80 | 8.49 | 201.79 | 69.64 | 8.41 | 202.19 |
| | Solution time of MP (s) | —— | 50.46 | 1119.45 | 5.89 | 5.89 | 66.89 | 5.99 | 5.99 |
| | Solution time of sub-problerm (s) | 0.28 | 6.12 | 2.35 | 2.60 | 195.90 | 2.75 | 2.42 | 196.20 |
| | Number of MGs | —— | —— | —— | 12 | 12 | —— | 12 | 12 |
| 45-24 | Objective function value | 131598.61 | 131595.42 | 131595.42 | 131595.42 | 131596.87 | 131595.42 | 131595.42 | 131596.87 |
| | Number of iterations | —— | 8 | 4 | 3 | 3 | 4 | 3 | 3 |
| | Solution time (s) | 0.35 | 111.6 | 2207.57 | 9.99 | 202.03 | 95.78 | 9.14 | 201.55 |
| | Solution time of MP (s) | —— | 103.23 | 2204.39 | 6.37 | 6.37 | 92.01 | 5.98 | 5.98 |
| | Solution time of sub-problerm (s) | 0.35 | 8.37 | 3.18 | 3.62 | 195.66 | 3.77 | 3.16 | 195.57 |
| | Number of MGs | —— | —— | —— | 15 | 15 | —— | 15 | 15 |
| 63-24 | Objective function value | 184236.69 | 184233.58 | 184233.58 | 184233.58 | 184260.95 | 184233.58 | 184233.58 | 184260.95 |
| | Number of iterations | —— | 9 | 5 | 3 | 3 | 5 | 3 | 3 |
| | Solution time (s) | 0.37 | 386.24 | 7949.41 | 12.96 | 178.82 | 261.14 | 10.7 | 176.55 |
| | Solution time of MP (s) | —— | 371.55 | 7944.56 | 8.12 | 8.12 | 256.21 | 5.99 | 5.99 |
| | Solution time of sub-problerm (s) | 0.37 | 14.69 | 4.85 | 4.84 | 170.70 | 4.93 | 4.71 | 170.56 |
| | Number of MGs | —— | —— | —— | 21 | 21 | —— | 21 | 21 |

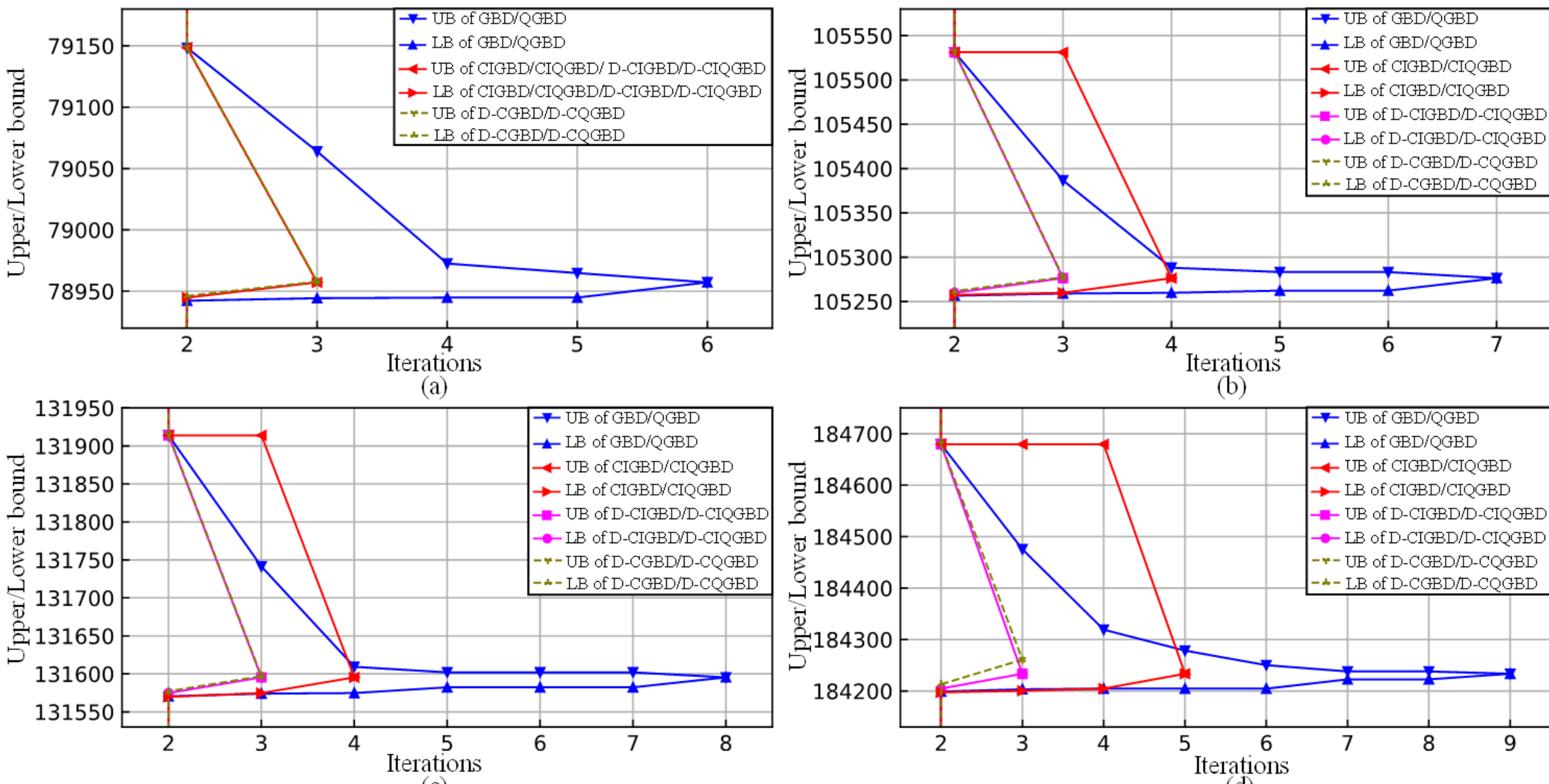

**Figure 8.** The UBs and LBs of different algorithms for UC problems with (a) 27 DERs, (b) 36 DERs, (c) 45 DERs, and (d) 63 DERs.

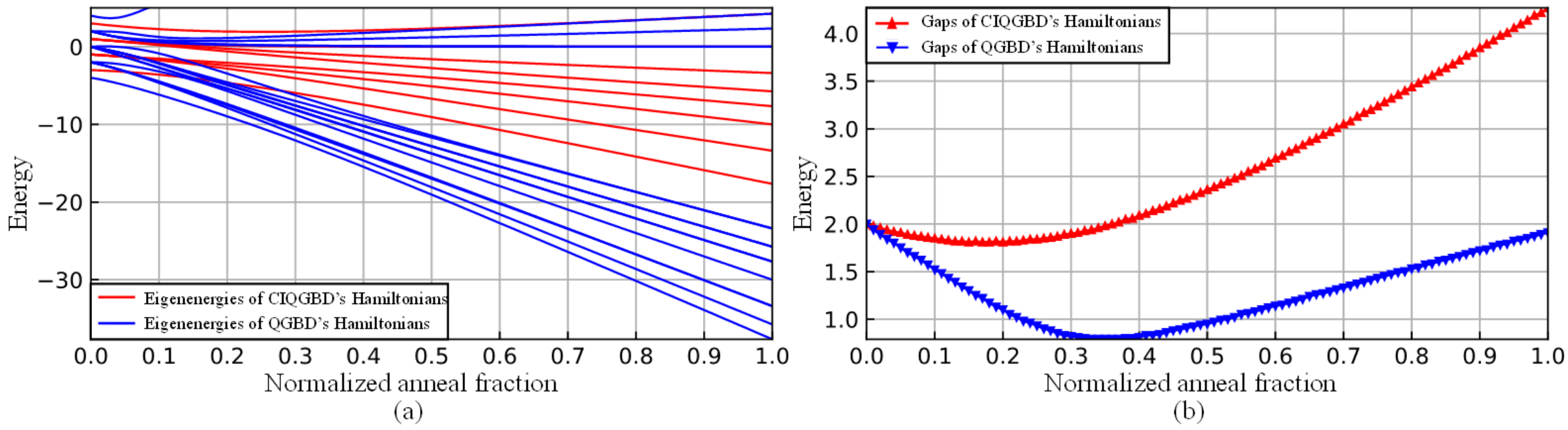

**Figure 9.** (a) The eigenenergy level distribution, and (b) the eigenenergy gap between the ground state and the first excited state of the system Hamiltonian during simulated quantum annealing process of CIQGBD and QGBD.

With the increase of the proportion of the problem Hamiltonian, the first excited state of the system gets closer to the ground state and then diverges. The minimum energy gap in QGBD is 0.79, which is much smaller than that in CIQGBD (*i.e.,* 1.81). Therefore, the problem Hamiltonian or the QUBO-MP constructed by CIQGBD is more reasonable than QGBD, which benefits from the more reasonable feasibility Benders cutting planes generated by the consensus-inspired feasible sub-problem and less introduced auxiliary binary variables. The less reasonable construction of the problem Hamiltonian in QGBD leads to the smaller minimum energy gap in the process of quantum annealing. As a result, when the number of DERs increases to 18, the system does not stay in the ground state after the annealing process, which means that QGBD can only get a suboptimal solution in this case. In distributed method D-CIQGBD, the order of the problem Hamiltonian mapped from the QUBO-LMP is further reduced, which better ensures that the system stays in the ground state in the process of quantum annealing.

*5.2.4. Scenario 5*

Compared with the centralized framework, the distributed framework can deal with larger UC problems. To investigate the performance improvement of D-CIQGBD compared with its classical version D-CIGBD and centralized methods (CIGBD, CIQGBD), the number of DERs is gradually increased to 99 in this scenario. The UC problem only considers the physical constraints of the unit.

**Figure 10** (a) compares the solution time of CIGBD, CIQGBD, D-CIGBD, and D-CIQGBD for UC problems at different scales, and shows that quantum algorithms CIQGBD and D-CIQGBD are faster than their corresponding classical versions CIGBD and D-CIGBD, respectively. The distributed algorithms D-CIGBD and D-CIQGBD exhibit significantly shorter solution times compared to the centralized algorithms CIGBD and CIQGBD. Moreover, this advantage becomes even more pronounced as the number of DERs increases. **Figure 10** (b) shows the solution time of D-CIGBD and D-CIQGBD for the LMP/sub-problem of UC problems with different total number of DERs and the same number (*i.e.*, 3) of DERs in each MG. The quantum version D-CIQGBD demonstrates a progressively more evident speed advantage over the classical version D-CIGBD as the number of DERs allocated within each MG increases. For instance, assuming a total of 90 DERs are distributed, with each MG containing $N_{DER}$ DERs, **Figure 10** (c) and (d) compare the total solution time, as well as the solution times for respective sub-problems and LMPs, for different values of $N_{DER}$ using D-CIQGBD and D-CIGBD. When the number of DERs allocated within each MG increases to three or more, the advantages of utilizing the quantum annealer D-WAVE for solving combinatorial optimization problems become evident. This results in significantly faster solving times for both LMPs and the overall solution time in D-CIQGBD compared to D-CIGBD.

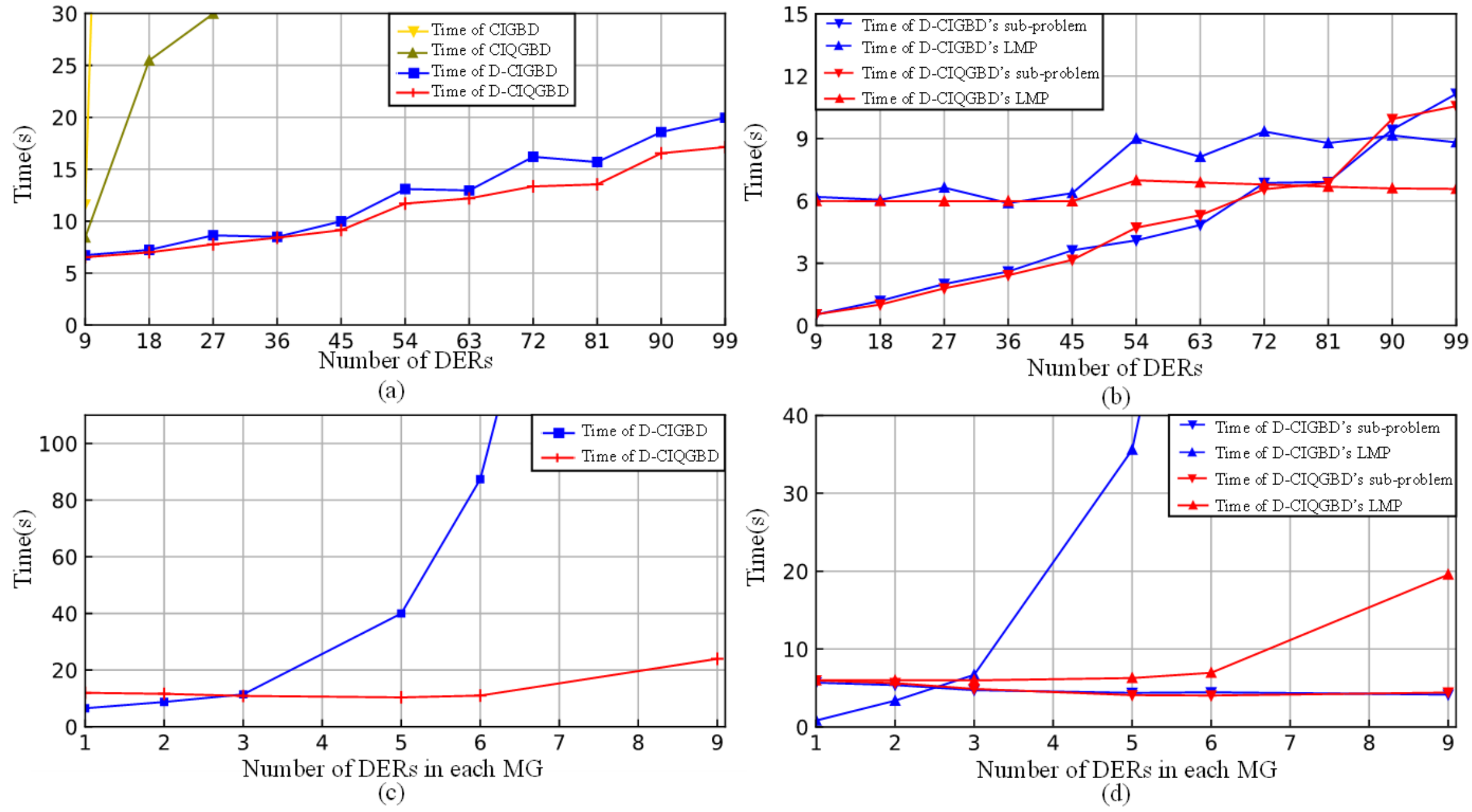

**Figure 10.** (a)(b) The solution time of CIGBD, CIQGBD, D-CIGBD, and D-CIQGBD for distributed UC problems with different total numbers of DERs (3 DERs in each MG); (c)(d) The solution time of D-CIGBD and D-CIQGBD for distributed UC problems with different numbers of DERs in each MG and 90 DERs in total.

*5.2.5. Scenario 6*

To further investigate the quantum advantage of the proposed hybrid quantum-classical algorithms, the quantum algorithms with the shortest solution time when solving a relatively large-scale UC problem in Scenario 4 (*i.e.*, CIQGBD and D-CIQGBD) are used to solve the SCUC problem, which extends the original UC problem by incorporating bus voltage phase angle variables and transmission line power flow variables. SCUC includes system power flow constraints and transmission line power flow capacity limits, while also transforming the overall power balance constraint into bus-level power balance constraints. While these modifications increase the complexity of the model, they also enhance its relevance to real-world applications.

In this scenario, we begin by applying the centralized algorithm CIQGBD to the 6-Bus System to assess the viability of the consensus-inspired strategy in addressing the SCUC problem. After confirming its effectiveness, we apply D-CIQGBD to solve the more complex IEEE RTS 24-Bus System. The 6-Bus System and the load variation are shown in **Figure 11** [32], while the IEEE RTS 24-Bus System and the load variation are shown in **Figure 12** [33]. **Table 4** presents the results of these two hybrid quantum-classical algorithms when applied to the 6-Bus System and IEEE RTS 24-Bus System. The solutions provided by CIQGBD and D-CIQGBD are consistent with those from the commercial solver Gurobi, and when the experimental scenario is IEEE RTS 24-Bus System, the solution time of D-CIQGBD outperforms that of Gurobi.

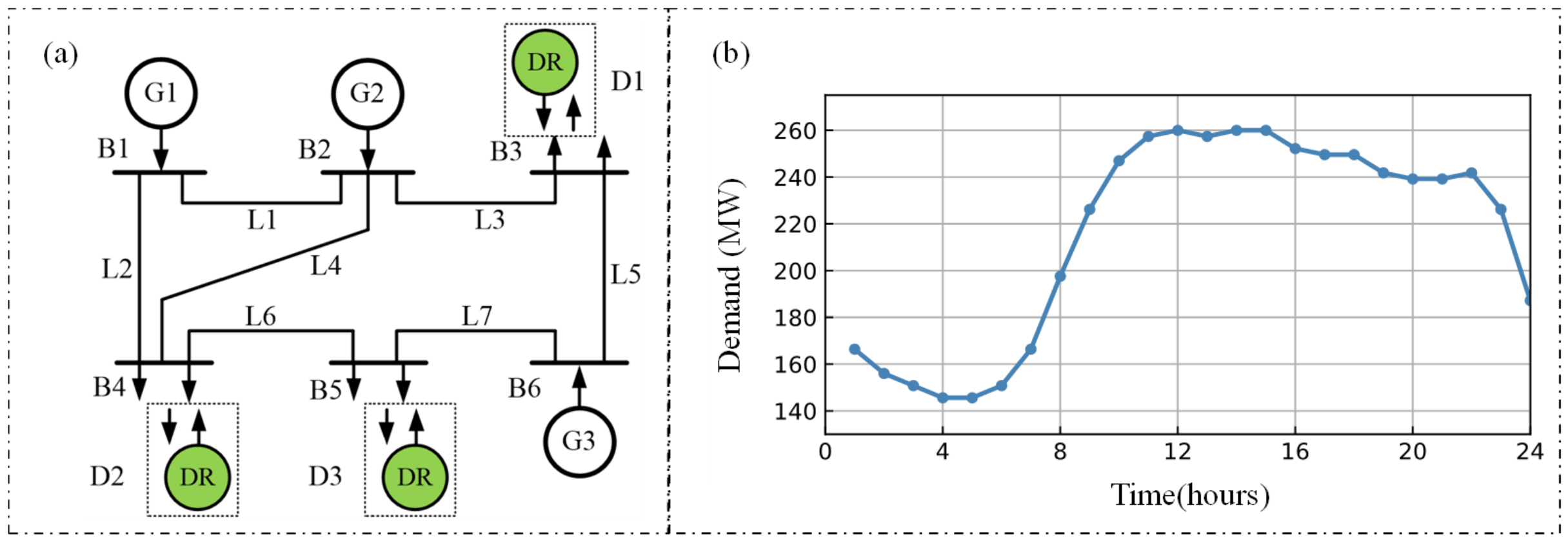

**Figure 11.** (a) Schematic diagram, and (b) Load variation of the 6-Bus System.

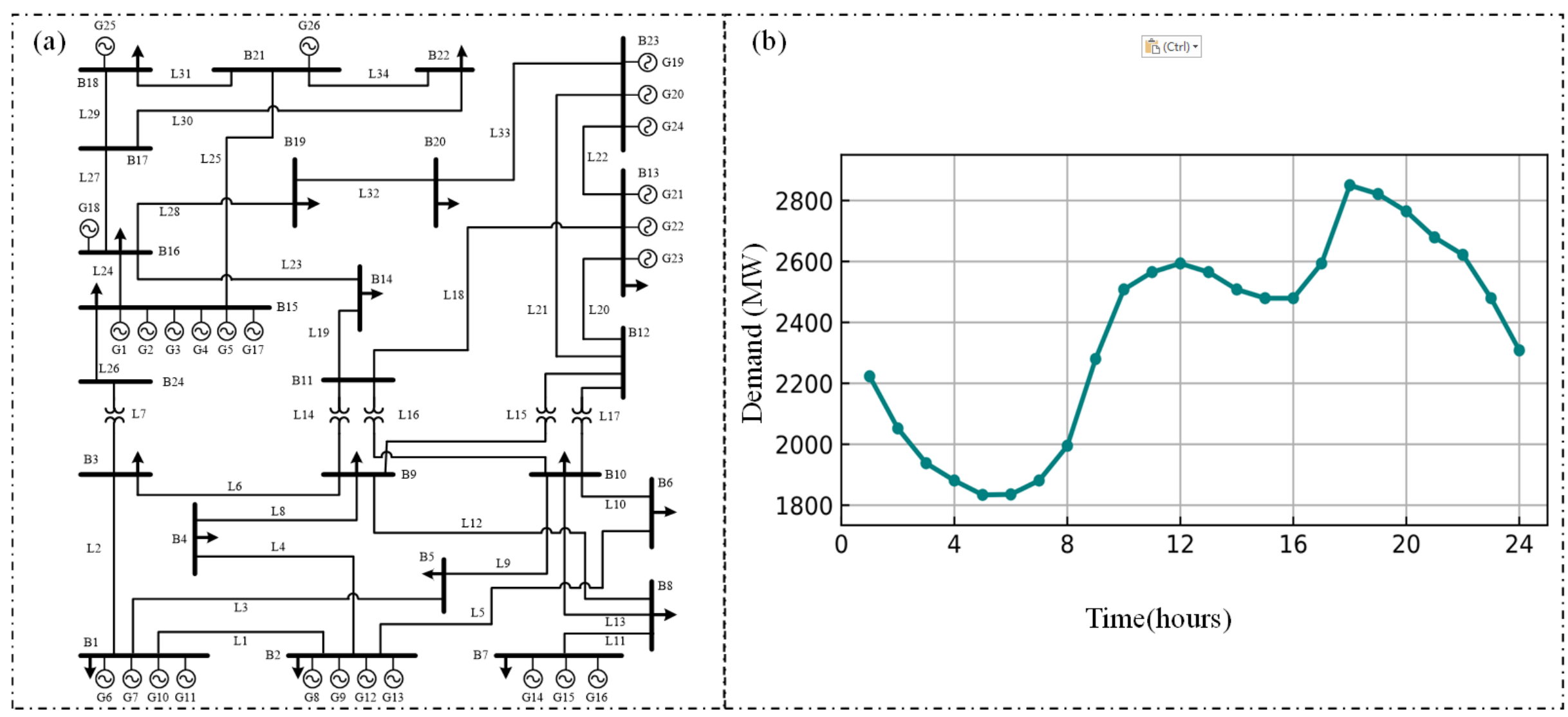

**Figure 12.** (a) Schematic diagram, and (b) Load variation of the IEEE RTS 24-Bus System.

**Table 4.** Performance comparison of Gurobi, CIQGBD and D-CIQGBD

| System | Data | Gurobi9 | CIQGBD | D-CIQGBD |
|---|---|---|---|---|
| 6-Bus System | Objective function value | 102547.60 | 102547.60 | |
| | Number of iterations | —— | 3 | |
| | Solution time (s) | 0.0450 | 0.3270 | —— |
| | Solution time of QPU (s) | —— | 0.3180 | |
| | Solution time of CPU (s) | 0.0450 | 0.0090 | |
| IEEE RTS 24-Bus System | Objective function value | 738159.68 | | 738159.68 |
| | Number of iterations | —— | | 5 |
| | Solution time (s) | 10.3473 | —— | 8.2175 |
| | Solution time of QPU (s) | —— | | 7.3383 |
| | Solution time of CPU (s) | 10.3473 | | 0.8792 |

## 6. Conclusions

For problems of centralized and distributed unit commitment, a series of hybrid quantum-classical generalized Benders decomposition algorithms are proposed in this work to make use of the advantage of quantum computing in solving QUBO problems [34]. The hybrid algorithms exhibit quantum advantage in terms of solving speed when tackling larger-scale UC problems compared to their classical counterparts.

In the centralized framework, we propose a quantum version of GBD (*i.e.,* QGBD), where the QUBO-MP can be solved using quantum algorithms, thereby significantly accelerating the computation process. In the distributed framework, D-CGBD and D-CQGBD introduce consensus variables between neighboring MGs, leading to the reformulation of the sub-problems as local sub-problems within each MG. Through consensus iterations and the convergence of these consensus variables, decoupled local cutting planes among MGs are obtained. Consequently, the LMP and QUBO-LMP are reconfigured. These two fully distributed algorithms (*i.e.*, D-CGBD and D-CQGBD) solve the UC problem via achieving consensus with neighboring MGs, ensuring that no MG discloses its internal private information. At the same time, these algorithms reduce the qubit requirements and meet the requirements for independent decision-making across regions and hierarchical partitioning in power grid management.

However, in D-CGBD and D-CQGBD, each MG must repeatedly communicate and reconcile its state with its neighboring microgrids to achieve consensus, and this process is inherently time-consuming. Drawing inspiration from the consensus among units, CIQGBD allocates slack variables more reasonably based on unit cost information. This approach reconstructs the relaxed feasible sub-problem, thereby enhancing the optimization of Benders cutting planes in the MP. By intuitively constructing the problem Hamiltonian, CIQGBD results in a more reasonable approach and a smaller minimum energy gap. This advantage proves beneficial in the quantum annealing process. This consensus-inspired strategy enables a more rational allocation of relaxation variables, allowing feasible subproblems to achieve optimal relaxation variable solutions under global optimization in a single step. This eliminates the computational burden of repeatedly adjusting consensus variables, significantly reducing subproblem solution time. Moreover, the improved allocation of relaxation variables ensures that CIQGBD naturally achieves variable consensus between neighboring microgrids (MGs), thereby facilitating the construction of local cutting planes. Therefore, CIQGBD seamlessly extends to the partially distributed framework D-CIQGBD. In the more practical context of SCUC, the proposed partially distributed algorithm, D-CIQGBD, demonstrates a significantly faster solution speed compared to the solver Gurobi.

## Acknowledgements

This work was supported by the National Key Research and Development Program of China under Grant. 2022YFB3304700, the Open Research Project of the State Key Laboratory of Industrial Control Technology under Grant. ICT2024B21, and the National Natural Science Foundation of China under Grant. 62273016.